\documentclass[12pt,eqsecnum]{article}
\usepackage[dvips]{graphicx}
\usepackage{amssymb}
\usepackage{amsmath}
\usepackage{epstopdf}
\usepackage{color}
\usepackage{soul}
\usepackage{cite}
\makeatletter

\@addtoreset{equation}{section}
\makeatletter

\newcommand{\qed}{\hbox{\rule[-2pt]{6pt}{6pt}}}
\newcommand{\D}{{\rm d}}

\newtheorem{lm}{Lemma}

\newcommand{\dalm}{\kern1pt\vbox{\hrule height 0.9pt\hbox{\vrule width
0.9pt\hskip 2.5pt\vbox{\vskip 5.5pt}\hskip 3pt\vrule width 0.3pt}\hrule height
0.3pt}\kern1pt}

\begin{document}

\begin{titlepage}
\vfill
\begin{flushright}
\today
\end{flushright}

\vfill
%\vskip 1.0cm
\begin{center}
\baselineskip=16pt
{\Large\bf 
Exact black-hole formation with a conformally coupled scalar field in three dimensions\\
}
\vskip 0.5cm
{\large {\sl }}
\vskip 10.mm
{\bf  Luis Avil\'es${}^{a, b}$, Hideki Maeda${}^{c}$, and Cristi{\'a}n Mart\'{\i}nez$^{a}$} \\

\vskip 1cm
{
    ${}^a$ Centro de Estudios Cient\'{\i}ficos (CECs), Av. Arturo Prat 514, Valdivia, Chile. \\
     ${}^b$ Departamento de F\'{\i}sica, Universidad de Concepci\'on, Casilla 160-C, Concepci\'on, Chile. \\
   	${}^c$ Department of Electronics and Information Engineering, Hokkai-Gakuen University, Sapporo 062-8605, Japan.\\
	\texttt{aviles@cecs.cl, h-maeda@hgu.jp, martinez@cecs.cl}

     }
\vspace{6pt}
%\today
\end{center}
\vskip 0.2in
\par
\begin{center}
{\bf Abstract}
\end{center}
\begin{quote}
We present exact dynamical and inhomogeneous solutions in three-dimensional AdS gravity with a conformally coupled scalar field.
They contain stealth configurations of the scalar field overflying the BTZ spacetime and also solutions with a non-vanishing energy-momentum tensor.
The latter non-stealth class consists of the solution obtained by Xu and its analytic extension.
It is shown that this proper extension represents: (i) an eternally shrinking dynamical black hole, (ii) a curious spacetime which admits an event horizon without any trapped surface, or (iii) gravitational collapse of a scalar field in an asymptotically AdS spacetime.
In the last case, by attaching the solution regularly to the past massless BTZ spacetime with a vanishing scalar field, the whole spacetime represents the black-hole formation from regular initial data in an asymptotically AdS spacetime.
Depending on the parameters, the formed black hole can be asymptotically static in far future.
\vfill
% \hrule width 5.cm
\vskip 2.mm
\end{quote}
\end{titlepage}

%<<<<<<<<<<<<< PACS NUMBER >>>>>>>>>>>>>>>%
%\pacs{
%04.50.-h 	Higher-dimensional gravity and other theories of gravity
%04.50.Gh 	Higher-dimensional black holes, black strings, and related objects 
%04.60.-m 	Quantum gravity
%04.60.Ds 	Canonical quantization 
%04.60.Kz 	Lower dimensional models; minisuperspace models 
%} 

%\maketitle
%\section{}
%\subsection{}

\tableofcontents

\newpage

%======================================%
%<<<<<<<<<<<< SECTION I  >>>>>>>>>>>>>>%
%======================================%
\section{Introduction}
\label{introduction}
In the last century, asymptotically anti-de~Sitter (AdS) spacetimes had been mostly ignored for a long time in the research of general relativity because isolated objects, like stars or black holes in our universe,  should be well approximated by asymptotically flat spacetimes. Moreover, cosmological observations prefer a positive value rather than negative for a cosmological constant.
However, at the end of the last century, there appeared several different and intriguing results associated with asymptotically AdS spacetimes, which triggered an explosive research trend of AdS gravity in the community. 
They are, for instance, the discovery of an unexpected three-dimensional black-hole spacetime which is obtained by identifications in the AdS spacetime~\cite{BTZ,BHTZ}, the possibility of non-spherical horizon topology of a black hole~\cite{topological}, and the AdS/CFT duality~\cite{ads/cft}. Especially, the last one is a duality between an asymptotically AdS spacetime and a conformal field theory (CFT) found by Maldacena, where AdS black holes are important tools to explore the properties of gauge theories in the strong coupling region. 
As a result, the extensive work in AdS gravity in the last decades has shown the physical and mathematical rich structure of the asymptotically AdS spacetimes, as the precursory works~\cite{Henneaux:1985tv,Brown:1986nw} early revealed in the 80's.

An example of such a fertile structure in AdS gravity is the configuration of the static and spherically symmetric black holes with scalar fields.
Indeed, higher-dimensional unified theories naturally predict scalar fields with a variety of potentials depending on the way of dimensional reduction.
However, in the case of asymptotically flat spacetimes, it is well known that black holes do not allow nontrivial configurations of scalar fields for positive convex potentials.
(See~\cite{Bekenstein:1995un} for instance and references therein.)
This result can be generalized for asymptotically dS spacetimes, namely in the presence of a positive cosmological constant \cite{ba2007}.
Actually, in four spacetime dimensions for certain types of potential or with a non-minimal coupling to the scalar curvature, asymptotically flat~\cite{Bocharova:1970skc,Bekenstein:1974sf}, dS \cite{Martinez:2002ru,Zloshchastiev:2004ny} or AdS~\cite{Martinez:2004nb,Martinez:2005di} black holes can admit nontrivial configurations of a scalar field, namely a {\it scalar hair}. The discovery of such a class of hairy black holes have stimulated an intensive search for exact solutions in AdS gravity
in presence of a real self-interacting scalar field, non or minimally coupled to Ricci scalar, in the last decade. (See  for instance  
\cite{Nadalini:2007qi, Martinez:2009kua,Kolyvaris:2009pc,Anabalon:2012tu,Anabalon:2012ta,Gonzalez:2013aca,Bardoux:2013swa,Acena:2013jya,Anabalon:2013eaa,Astorino:2014mda,Fan:2015tua,Ayon-Beato:2015ada}.) 
However, it is known that some of the asymptotically flat or dS scalar hairy black holes are dynamically unstable against spherical perturbations~\cite{Bronnikov:1978mx,Harper:2003wt,Dotti:2007cp}.
In contrast, it has been shown that some asymptotically AdS black holes with scalar hair, obtained by numerical methods, are dynamically stable \cite{tmn2001,winstanley2005}. (See  also  \cite{Anabalon:2013baa} for other results.)

Also, nonlinear instability of the AdS spacetime~\cite{br2011,dhs2012} expresses a sharp difference from the Minkowski and dS spacetimes.
It is known that the Minkoswki spacetime is nonlinearly stable~\cite{flat-stability} and the dS spacetime is stable against small perturbations~\cite{dS-stability}.
But in contrast, the AdS spacetime suffers from the turbulent instability and finally results in the formation of a curvature singularity by gravitational collapse.
While it has been claimed that many asymptotically AdS spacetimes are nonlinearly stable~\cite{dhms2012}, the final state of the gravitational collapse caused by this turbulent instability has not been clarified yet.

In the AdS/CFT context, formations of a black hole from gravitational collapse or evolving black holes correspond to CFTs describing non-equilibrium states at the boundary.
Therefore, such dynamical AdS spacetimes containing a black hole could provide us a chance to study such non-equilibrium states in condensed matter physics which are not well-understood at present.
In this context, dynamical AdS black holes~\cite{kmno,bbl2017} or formations of an AdS black hole~\cite{bll2012} have been certainly studied, however, most of the solutions were constructed numerically.
Of course, in order to derive more specific results in an analytic manner, exact solutions are desirable.
The Vaidya-AdS solution for a null dust fluid~\cite{vaidya,GP-text} and the Lema\^{\i}tre-Tolman-Bondi-AdS solution for a timelike dust fluid~\cite{LemaitreTolmanBondi,GP-text} are examples of such exact solutions.
(See also~\cite{Zhang:2014sta,Zhang:2014dfa,Fan:2015tua,Fan:2015ykb} for recent examples.)
Since the original AdS/CFT duality was found and its generalization, namely the AdS/CFT conjecture, was proposed in the context superstring/M-theory, solutions with fundamental fields such as a scalar field or a gauge field must be suitable as a holographic dual to the CFT at the AdS boundary.
A main subject of the present paper is to provide such a class of solutions.

For this purpose, we focus on AdS black holes with a conformally coupled scalar field in three dimensions in the present work.
With this class of a non-minimally coupled scalar field, a 
static hairy black hole was obtained twenty years ago~\cite{Martinez:1996gn, Henneaux:2002wm}.
Subsequently, the so-called \textit{stealth} configurations of scalar fields overflying the BTZ black hole, characterized by a vanishing energy-momentum tensor, were found~\cite{AyonBeato:2004ig}. Further static \cite{Correa:2011dt} and rotating \cite{Natsuume:1999at,Correa:2012rc,Liu:2014eqa} hairy three-dimensional black holes were built by adding suitable potentials. More recently, Xu obtained an exact dynamical and inhomogeneous solution which represents gravitational collapse~\cite{Xu:2014xqa}. 

In this article, we consider a circularly symmetric spacetime with a conformally coupled real scalar field, depending  on time and the radial coordinate, in the presence of a negative cosmological constant. 
Then, we present all the possible solutions of the fields equations, which contains not only the known solutions mentioned before but also new ones.  
Among them, in  particular, there is a solution representing gravitational collapse of a scalar field in an asymptotically AdS spacetime.
Finally, attaching this solution regularly to the past massless BTZ spacetime with a vanishing scalar field, we construct a maximally extended spacetime which represents formation of an AdS black hole from regular initial data.

In the next section, we first present the action and the field equations of the system. 
After deriving the curvature tensors and geodesic equations in the spacetime we consider, the concept of trapping horizon is briefly reviewed.
Subsequently, we will demonstrate for the readers an analysis how to identify the final state of gravitational collapse in the case of the Vaidya-AdS solution for a null dust fluid.
This is a useful practice providing a methodology for the main result obtained in the following sections. 
In Sec.~\ref{sec:exact}, we will present all the possible solutions under our metric assumption, which contain not only an analytic extension of the Xu's non-stealth solution but also new configurations of a stealth scalar field overflying the BTZ spacetime. In the original paper by Xu, some properties of a non-stealth solution have been studied, but it is difficult to grab them in a transparent manner because of the complicated form of the metric coming from an unsuitable choice of coordinates.
In the present article, we will show that the Xu's solution can be written in a much simpler form by coordinate transformations and then clearly expose its geometrical and physical properties.
Actually, in our coordinate system, Xu's solution is analytically extended into the ``hidden'' domain in the previous coordinate system and it is shown that there exists another branch of solutions which represents a distinct spacetime.
In Sec.~\ref{sec:main}, by a careful analysis of the geometrical and physical properties, we will show that the extended non-stealth solution describes a variety of physically interesting situations depending on the parameters.
Concluding remarks and future prospects are given in the final section.

Our basic notations follow~\cite{wald}.
The conventions of curvature tensors are 
$[\nabla _\rho ,\nabla_\sigma]V^\mu ={R^\mu }_{\nu\rho\sigma}V^\nu$ 
and ${R}_{\mu \nu }={R^\rho }_{\mu \rho \nu }$.
The Minkowski metric has the signature $(-,+,+)$ and Greek indices run over all spacetime indices.
We adopt the units such that $c=1$ and $\kappa$ denotes  the three-dimensional gravitational constant.

%======================================%
%<<<<<<<<<<<< SECTION I  >>>>>>>>>>>>>>%
%======================================%
\section{Preliminaries}
\subsection{Model and field equations}
In the present paper, we consider the action for gravity coupled to a non-minimally self-interacting scalar field $\phi$  in three spacetime dimensions in the presence of a negative cosmological constant $\Lambda$:
\begin{equation}
I[g_{\mu \nu}, \phi]=\int \D^{3}x \sqrt{-g}\left( \frac{R -2 \Lambda}{2\kappa}-\frac{1}{2}g^{\mu \nu}\partial_\mu \phi \partial_\nu \phi-\frac{\xi}{2}R \phi^2-\alpha \phi^6 \right),  \label{action-1000}
\end{equation}
where $\xi$ is a non-minimal coupling parameter and $\alpha$ is a coupling constant to the self-interaction potential $V(\phi)$. Hereafter, we set $\xi=1/8$ for ensuring the matter piece of the action is invariant under the conformal transformations $g_{\mu \nu}\rightarrow \Omega^2 g_{\mu \nu}$ and $\phi \rightarrow \Omega ^{-1/2} \phi$.

The field equations derived from the above action are
\begin{align}
G_{\mu \nu}-l^{-2} g_{\mu \nu} &= \kappa T_{\mu \nu}, \label{Einstein}\\ 
\square \phi - \frac{1}{8}R \phi -6\alpha \phi^5 &= 0,  \label{FE}
\end{align}
where $l$ is the AdS radius defined by $l^{-2}:=-\Lambda$ and the energy-momentum tensor for a conformally coupled scalar field is
given by
\begin{equation}
T_{\mu \nu}= \partial_\mu \phi \partial_\nu \phi -\frac{1}{2}g_{\mu \nu}g^{\alpha \beta}\partial_\alpha \phi \partial_\beta \phi -\alpha g_{\mu \nu}\phi^6+\frac{1}{8}(g_{\mu \nu} \square-\nabla_\mu \nabla_\nu+ G_{\mu \nu})\phi^2. 
\end{equation}
This energy-momentum tensor is traceless which is a characteristic of a conformally coupled field.
For the following sections it is convenient to define a constant $\beta$, which depends only of the parameters of the action, given by
\begin{equation}
\beta:= \frac{512 \alpha l^2-\kappa^2}{8 \kappa l^2}. \label{def-beta}
\end{equation}

\subsection{Metric assumption and geometric properties}
In the present paper, we will discuss spacetimes with circular symmetry described in the following coordinates $(v,r,\theta)$:
\begin{align}
\D s^2=-f(v,r)\D v^2+2\D v\D r+r^2\D\theta^2,\label{metric-assumption}
\end{align}
where $v$ is the advanced time, $r$ is the areal radius, and $\theta \in [0, 2\pi]$ is the angular coordinate.
We define the future direction by an increasing direction of $v$.

In what follows we summarize geometrical properties of the spacetime (\ref{metric-assumption}) for the later use. 
The nonzero components of the Christoffel symbol for the spacetime (\ref{metric-assumption}) are
\begin{align} 
\label{Gammas}
\begin{aligned}
&\Gamma^v_{vv}=\frac12f_{,r},\quad \Gamma^v_{\theta\theta}=-r,\quad \Gamma^r_{vv}=\frac12(ff_{,r}-f_{,v}), \\
&\Gamma^r_{vr}=-\frac12f_{,r},\quad \Gamma^r_{\theta\theta}=-rf,\quad \Gamma^\theta_{r\theta}=r^{-1},
\end{aligned}
\end{align}
where a comma denotes a partial derivative. From Eq.~\eqref{Gammas} we get the nonzero components of the Riemann tensor $R^{\mu\nu}_{~~\rho\sigma}$ and the Ricci tensor $R^{\mu}_{~\nu}$:
\begin{align}
\begin{aligned}
&R^{vr}_{~~vr}=-\frac12f_{,rr},\quad R^{v\theta}_{~~v\theta}=R^{r\theta}_{~~r\theta}=-\frac{1}{2r}f_{,r},\quad R^{r\theta}_{~~v\theta}=-\frac{1}{2r}f_{,v},\\
&R^v_{~v}=R^r_{~r}=-\frac{1}{2r}(rf_{,rr}+f_{,r}),\quad R^v_{~r}=-\frac{1}{2r}f_{,v},\quad R^\theta_{~\theta}=-\frac{1}{r}f_{,r}.
\end{aligned}
\end{align}
Lastly, the Ricci scalar $R$ and the Kretschmann scalar $K:=R_{\mu\nu\rho\sigma}R^{\mu\nu\rho\sigma}$ are given by 
\begin{align} \label{RandK}
R=-\frac{1}{r}(rf_{,rr}+2f_{,r}),\qquad  K=(f_{,rr})^2+\frac{2}{r^2}(f_{,r})^2.
\end{align}

\subsubsection{Geodesic equations}
Here we derive geodesic equations in the spacetime (\ref{metric-assumption}).
Let us consider an affinely-parametrized geodesic $x^\mu(\lambda)=(v(\lambda),r(\lambda),\theta(\lambda))$, where $\lambda$ is an affine parameter.
In this spacetime, $L:={\bar k}_\mu \xi^\mu_{(\theta)}=r^2{\dot \theta}$ is a conserved quantity along any geodesic, where ${\bar k}^\mu=({\dot v},{\dot r},{\dot \theta})$ is the tangent vector of the geodesic and a dot denotes differentiation with respect to $\lambda$.
This constant $L$ is associated with the Killing vector $\xi_{(\theta)}^\mu=(0,0,1)$ generating a circular symmetry and interpreted as the angular momentum of the particle moving along the geodesic.

Using $L$, we can write the first integral along the geodesic as
\begin{align}
\varepsilon=&-f{\dot v}^2+2{\dot v}{\dot r}+\frac{L^2}{r^2},\label{1stintegral}
\end{align}
where $\varepsilon=-1,0,+1$ for timelike, null, and spacelike geodesics, respectively.
With a help of the first integral (\ref{1stintegral}), geodesic equations ${\ddot x}^\mu+\Gamma^\mu_{\rho\sigma}{\dot
x}^\rho{\dot x}^\sigma=0$ are explicitly written as
\begin{align}
0=&{\ddot v}+\frac12f_{,r}{\dot v}^2-\frac{L^2}{r^3},\\
0=&{\ddot r}-\frac12f_{,v}{\dot v}^2-\frac12\varepsilon f_{,r}+\frac{L^2}{2r^3}(rf_{,r}-2f).
\end{align}

Equation~(\ref{1stintegral}) shows that 
\begin{align}
{\dot v}(f{\dot v}-2{\dot r})\ge 0
\end{align}
is satisfied for causal geodesics with equality holding for radial null geodesics ($\varepsilon=L=0$).
We can set ${\dot v}>0$ without loss of generality if $v(\lambda)$ is not a constant function, with which $\D r/\D v\le f/2$ holds along such causal geodesics.
Then, constant $v$ with decreasing $r$ represents future-directed radial {\it ingoing} null geodesics, while future-directed radial {\it outgoing} null geodesics satisfy
\begin{align}
\frac{\D v}{\D r}=\frac{2}{f},\label{ODE-radialnull}
\end{align}
which plays a crucial role in the analysis of the global structure of the spacetime.

Actually, in order to understand the structure of the singularity if there is, we need to prove or disprove the existence of the solution for the geodesic equation (\ref{ODE-radialnull}).
The Lipschitz continuity is a well-known condition to prove the existence of the solution for an ordinary differential equation, such as the one given by (\ref{ODE-radialnull}).
Consider two points $(r,v_1)$ and $(r,v_2)$ in the $(r,v)$-plane and also a continuous function $w(r,v)$ in a given domain ${\cal D}$.
The Lipschitz condition for $w$ in ${\cal D}$ is that there exists a positive constant $\zeta$ such that 
\begin{align}
|w(r,v_1)-w(r,v_2)|<\zeta|v_1-v_2|
\end{align}
holds.
If $w$ satisfies the Lipschitz condition, there is a unique $C^1$ solution in ${\cal D}$ for the ordinary differential equation $\D v/\D r=w(r,v)$ with a given initial condition $(r_0,v(r_0))\in {\cal D}$.
It is noted that if there exits a derivative $\partial w/\partial v$ which is continuous and bounded in ${\cal D}$, then the Lipschitz condition is satisfied by the mean value theorem.
For the future-directed radial outgoing null geodesic equation (\ref{ODE-radialnull}), we have 
\begin{align}
w=\frac{2}{f},\qquad w_{,v}=-\frac{2f_{,v}}{f^2}. \label{L-function}
\end{align}

\subsubsection{Orthonormal bases}
Along a radial null geodesic in the spacetime (\ref{metric-assumption}),
\begin{equation}
0=\D v(-f\D v+2\D r)
\end{equation}
is satisfied.
Ingoing radial null geodesics are represented by $v=$constant, while
outgoing null geodesics satisfy
\begin{equation} \label{out-null}
\frac{\D r}{\D v}=\frac12f.
\end{equation}
The tangent vectors along the future-directed radial outgoing and ingoing null geodesics, which are denoted respectively as $k^\mu$ and $l^\mu$, are given by 
\begin{equation}
k^\mu\frac{\partial}{\partial x^\mu}=\frac{\partial}{\partial v}+\frac{f}{2}\frac{\partial}{\partial r},\qquad l^\mu\frac{\partial}{\partial x^\mu}=-\frac{\partial}{\partial r}, \label{tangent-geodesics}
\end{equation}
which satisfy $k^\mu k_\mu=l^\mu l_\mu=0$ and $k^\mu l_\mu=-1$.

In addition to $k^\mu$ and $l^\mu$ defined by Eq.~(\ref{tangent-geodesics}), we define a unit spacelike vector $m^\mu$ given by  
\begin{equation}
m^\mu\frac{\partial}{\partial x^\mu}=\frac{1}{r}\frac{\partial}{\partial \theta},
\end{equation}
which satisfies $m_\mu k^\mu=m_\mu l^\mu=0$ and $m_\mu m^\mu=1$.
A set of three vectors $(k^\mu,l^\mu,m^\mu)$ forms a pseudo-orthonormal basis in the spacetime (\ref{metric-assumption}): 
\begin{equation}
{\bar E}^\mu_{(a)}=({\bar E}^\mu_{(0)},{\bar E}^\mu_{(1)},{\bar E}^\mu_{(2)})=(k^\mu,l^\mu,m^\mu).\label{psude-bases}
\end{equation}
This basis satisfies  
\begin{equation}
{\bar E}^\mu_{(a)}{\bar E}_{(b)\mu}={\bar \eta}_{(a)(b)}=\left(
\begin{array}{ccc}
 0  & -1 & 0 \\
 -1 & 0 & 0 \\
 0 & 0 & 1 \\
\end{array}
\right).
\end{equation}
Here ${\bar \eta}_{(a)(b)}$ is the metric in the local Lorentz frame and the metric $g_{\mu\nu}$ in the spacetime is given by $g_{\mu\nu}={\bar \eta}_{(a)(b)}{\bar E}^{(a)}_{\mu}{\bar E}^{(b)}_{\nu}$.
The basis ${\bar E}^\mu_{(a)}$ is parallelly transported along a null curve $v=$constant, namely $l^\nu\nabla_\nu {\bar E}^\mu_{(a)}=0$ holds.
The components of the Riemann tensor in the parallelly propagated pseudo-orthonormal frame (\ref{psude-bases}) are given by 
\begin{equation}
{\bar R}_{(a)(b)(c)(d)}:=R_{\mu\nu\rho\sigma}{\bar E}^\mu_{(a)}{\bar E}^\nu_{(b)}{\bar E}^\rho_{(c)}{\bar E}^\sigma_{(d)},
\end{equation}
of which nonzero components are
\begin{align}
{\bar R}_{(0)(1)(0)(1)}=&\frac{1}{2}f_{,rr},\quad {\bar R}_{(0)(2)(0)(2)}=-\frac{1}{2r}f_{,v},\quad {\bar R}_{(0)(2)(1)(2)}=\frac{1}{2r}f_{,r}.\label{R-psude}
\end{align}

Also, we construct a unit timelike vector $u^\mu$ and a unit spacelike vectors $s^\mu$ such that 
\begin{align}
u^\mu\frac{\partial}{\partial x^\mu}:=&\frac{1}{\sqrt{2}}\biggl(k^\mu\frac{\partial}{\partial x^\mu}+l^\mu\frac{\partial}{\partial x^\mu}\biggl)=\frac{1}{\sqrt{2}}\biggl\{\frac{\partial}{\partial v}+\biggl(\frac{f}{2}-1\biggl)\frac{\partial}{\partial r}\biggl\},\\
s^\mu\frac{\partial}{\partial x^\mu}:=&\frac{1}{\sqrt{2}}\biggl(-k^\mu\frac{\partial}{\partial x^\mu}+l^\mu\frac{\partial}{\partial x^\mu}\biggl)=\frac{1}{\sqrt{2}}\biggl\{-\frac{\partial}{\partial v}-\biggl(\frac{f}{2}+1\biggl)\frac{\partial}{\partial r}\biggl\},
\end{align}
which satisfy $u^\mu u_\mu=-1$, $s^\mu s_\mu=1$, $u^\mu s_\mu=0$, and $m_\mu u^\mu=m_\mu s^\mu=0$.
A set of three vectors $(u^\mu,s^\mu,m^\mu)$ forms parallelly transported orthonormal basis in the spacetime (\ref{metric-assumption}): 
\begin{equation}
{E}^\mu_{(a)}=({E}^\mu_{(0)},{E}^\mu_{(1)},{E}^\mu_{(2)})=(u^\mu,s^\mu,m^\mu).\label{normal-bases}
\end{equation}
This new basis verifies 
\begin{equation}
{E}^\mu_{(a)}{E}_{(b)\mu}={\eta}_{(a)(b)}=\left(
\begin{array}{ccc}
 -1  & 0 & 0 \\
 0 & 1 & 0 \\
 0 & 0 & 1 \\
\end{array}
\right)
\end{equation}
and the metric $g_{\mu\nu}$ is given by $g_{\mu\nu}={\eta}_{(a)(b)}{E}^{(a)}_{\mu}{E}^{(b)}_{\nu}$.
The components of the Riemann tensor in the parallelly propagated orthonormal frame (\ref{normal-bases}) are given by 
\begin{equation}
R_{(a)(b)(c)(d)}:=R_{\mu\nu\rho\sigma}E^\mu_{(a)}E^\nu_{(b)}E^\rho_{(c)}E^\sigma_{(d)},
\end{equation}
of which nonzero components are
\begin{align}
\label{R-ortho}
\begin{aligned}
R_{(0)(1)(0)(1)}=&\frac{1}{2}f_{,rr},\quad R_{(0)(2)(0)(2)}=\frac{1}{4r}(2f_{,r}-f_{,v}),\\
R_{(0)(2)(1)(2)}=&\frac{1}{4r}f_{,v}, \quad R_{(1)(2)(1)(2)}=-\frac{1}{4r}(2f_{,r}+f_{,v}).
\end{aligned}
\end{align}

\subsubsection{Trapping horizon}
Traditionally, a black hole is defined by the event horizon, which is a future boundary of the causal past of the future null infinity.
The event horizon is a global concept and one needs the information of the entire future of a spacetime to identify its location.
However, it is a difficult task.
For this reason, a quasi-local notion of horizon is often used in order to define a black-hole spacetime because its location is much easier to identify than the event horizon.

In stationary spacetimes, a Killing horizon associated with a Killing vector generating the symmetry of stationarity is a possible quasi-local definition of a black hole.
In general relativity, under certain physically reasonable assumptions, the rigidity theorem has been established asserting that the event horizon in a stationary spacetime is a Killing horizon~\cite{arealaw,Hawking:1973uf,Chrusciel:1996bj,Hollands:2006rj}.

However, we will consider dynamical spacetimes described by the metric (\ref{metric-assumption}), in which there is no timelike Killing vector.
In the case of such fully dynamical spacetimes, a trapping horizon has been proposed by Hayward as a quasi-local definition of a black hole~\cite{hayward1994,hayward1996}.
In the present paper, we also use a trapping horizon to identify a black-hole region in the spacetime.

Among all the classes of trapping horizons, a future outer trapping horizon defines a black hole. 
The idea of this definition is that the following three conditions hold on the horizon: (i) ingoing null rays should be converging, namely, their expansions $\Theta_-$ satisfy $\Theta_-< 0$, (ii) outgoing null rays should be instantaneously parallel, namely $\Theta_+= 0$, and (iii) outgoing null rays should be diverging outside the horizon and converging inside, namely ${\cal L}_-\Theta_+< 0$, where ${\cal L}_-$ is the Lie derivative along an ingoing null ray.

In general relativity, under the null energy condition, an outer trapping horizon is non-timelike, and the future domain of a future outer trapping horizon is a trapped region~\cite{hayward1994,hayward1996,nm2008}.
They clearly show that a future outer trapping horizon is a one-way membrane being matched to the concept of a black hole as a region of no escape.
In addition, in our spacetime (\ref{metric-assumption}), a future outer trapping horizon coincides with the Killing horizon in the static limit.

The location of a trapping horizon in the spacetime (\ref{metric-assumption}) is identified as follows.
Since we assumed that the domain of the angular coordinate is $0\le \theta\le
2\pi$, the surface area with constant $v$ and $r$ is given by
$\mathcal{A}:=2\pi r$.
Now expansions along outgoing and ingoing radial null geodesics are respectively computed as
\begin{align}
\Theta_+:=&\frac{k^\mu\nabla_\mu\mathcal{A}}{\mathcal{A}}=\frac{1}{\mathcal{A}}\biggl(\frac{\partial \mathcal{A}}{\partial
v}+\frac{f}{2}\frac{\partial \mathcal{A}}{\partial r}\biggl)
=\frac{1}{2r} f,\label{out-expansion}\\
\Theta_-:=&\frac{l^\mu\nabla_\mu\mathcal{A}}{\mathcal{A}}=-\frac{1}{\mathcal{A}}\frac{\partial \mathcal{A}}{\partial r}
=-\frac{1}{r},\label{expansion2}
\end{align}
where we used Eqs.~(\ref{out-null}) and~(\ref{tangent-geodesics}).

A trapping horizon is defined by the vanishing null expansion.
In the present case, since $\Theta_-$ is negative definite, the location of a trapping horizon $r=r_{\rm h}(v)$ is given by $\Theta_+=0$, namely it is obtained by solving the following algebraic equation:
\begin{align} \label{t-horizon}
f(r_{\rm h})=0.
\end{align}
A future outer trapping horizon is defined by $\Theta_+=0$ with $\Theta_-<0$ and ${\cal L}_-\Theta_+<0$.
While Eq.~(\ref{expansion2}) shows $\Theta_-<0$ in the present case, we compute
\begin{align} \label{der-expansion}
{\cal L}_-\Theta_+|_{r=r_{\rm h}}=&l^\mu \nabla_\mu \Theta_+|_{r=r_{\rm h}}=-\frac{1}{2r}\frac{\partial f}{\partial r}\biggl|_{r=r_{\rm h}},
\end{align}
where we used Eq.~(\ref{t-horizon}).
Thus, a future outer trapping horizon is realized for
\begin{align} \label{der-expansion-pos200}
\frac{\partial f}{\partial r}\biggl|_{r=r_{\rm h}}>0.
\end{align}
In summary, in the spacetime (\ref{metric-assumption}), the location of a future outer trapping horizon is determined by Eq.~(\ref{t-horizon}) with Eq.~(\ref{der-expansion-pos200}).

\subsection{Gravitational collapse of a null dust fluid: A practice}
\label{sec:practice}
An example of exact solutions in the form of the metric (\ref{metric-assumption}) is the three-dimensional Vaidya-AdS solution~\cite{husain1994}, which is a solution for a null dust fluid with negative $\Lambda$ in general relativity.
In this subsection, we review this solution and demonstrate an analysis how to identify the final state of gravitational collapse.
We will use most of the techniques in this subsection to obtain our main result in Sec.~\ref{sec:main}.

The energy-momentum tensor for a null dust fluid is given by 
\begin{align}
T_{\mu\nu}=\rho l_\mu l_\nu,
\end{align}
where $\rho$ is the energy density and $l^\mu$ is a null vector ($l^\mu l_\mu=0$).
The three-dimensional Vaidya-AdS solution is given by
\begin{align}
\D s^2=&-\biggl(\frac{r^2}{l^2}-\mu(v)\biggl)\D v^2+2\D v\D r+r^2\D\theta^2, \label{vaidya-gr}\\
l^\mu\frac{\partial}{\partial x^\mu}=&-\frac{\partial}{\partial r},\qquad \rho(v,r)=\frac{\mu'}{2\kappa r},\label{rho-gr}
\end{align}
where $\mu(v)$ is an arbitrary function and a prime denotes differentiation with respect to $v$.

The nonzero components of the Riemann tensor for the spacetime (\ref{vaidya-gr}) are given by 
\begin{align}
R^{vr}_{~~vr}=&R^{v\theta}_{~~v\theta}=R^{r\theta}_{~~r\theta}=-\frac{1}{l^2},\quad R^{r\theta}_{~~v\theta}=\frac{\mu'}{2r}.
\end{align}
In spite that all the curvature invariants are finite in this spacetime, there is a central curvature singularity at $r=0$ unless $\mu$ is constant.
Actually, this is not a scalar polynomial curvature singularity but a parallelly propagated (p.p.) curvature singularity, which is defined by the fact that some component of the Riemann tensor in the parallelly propagated frame blows up~\cite{Hawking:1973uf}.
Equation~(\ref{R-ortho}) shows that the following components in the parallelly propagated orthonormal frame along a null curve $v=$constant certainly blow up for $r\to 0$ unless $\mu$ is constant:
\begin{align}
R_{(0)(2)(0)(2)}=\frac{\mu'}{4r}+\frac{1}{l^2},\quad R_{(0)(2)(1)(2)}=-\frac{1}{4r}\mu', \quad R_{(1)(2)(1)(2)}=\frac{\mu'}{4r}-\frac{1}{l^2}.
\end{align}
Divergence is also observed in the following component in the parallelly propagated pseudo-orthonormal frame (\ref{R-psude}):
\begin{equation}
{\bar R}_{(0)(2)(0)(2)}=\frac{\mu'}{2r}. 
\end{equation}

We assume $\mu(v)=\mu_1v$ in the solution (\ref{vaidya-gr}), where $\mu_1$ is a positive constant because our non-stealth solution discussed in Sec. \ref{sec:main} obeys the same asymptotic behavior near $v=0$, of which parameter $B_0$ is related to $\mu_1$ as $\mu_1=2B_0^2/3\kappa (> 0)$.
With $\mu(v)=\mu_1 v$, the spacetime is asymptotically (at least) locally AdS for $r\to \infty$ along any curve and there is a future outer trapping horizon given by $v=r^2/(\mu_1l^2)$, which is spacelike.
The central singularity $r=0$ is in the trapped (untrapped) region in the domain of $v>(<)0$ because $f<(>)0$ is satisfied there.
As shown below, the singularity in the trapped (untrapped) region is spacelike (timelike).

We compute the functions in Eq.~(\ref{L-function}) as
\begin{align}
w=&2\biggl(\frac{r^2}{l^2}-\mu(v)\biggl)^{-1}=2\biggl(\frac{r^2}{l^2}-\mu_1 v\biggl)^{-1},\\
w_{,v}=&2\mu_{,v}\biggl(\frac{r^2}{l^2}-\mu(v)\biggl)^{-2}=2\mu_1\biggl(\frac{r^2}{l^2}-\mu_1v\biggl)^{-2}.
\end{align}
Since both $w$ and $w_{,v}$ are continuous and finite at and around the central singularity given by $(r,v)=(0,v_{\rm s})$ with $v_{\rm s}\ne 0$, there is a unique future-directed outgoing radial null geodesic $\gamma_{\rm out}$ satisfying $r(v_{\rm s})=0$.
In contrast, neither $w$ nor $w_{,v}$ is finite at $r=v=0$ and therefore, it is more subtle whether there exists a $\gamma_{\rm out}$ emanating from the central singularity at $r=v=0$, which will be clarified later.

Now let us clarify the signature of the central singularity with $v\ne 0$.
$v=v_{\rm s}$ represents a future-directed ingoing radial null geodesic $\gamma_{\rm in}$ terminating at the singularity at $(r,v)=(0,v_{\rm s})$.
If the singularity is located in the untrapped region ($f>0$), $r$ is a spacelike coordinate and so there is a single $\gamma_{\rm out}$ which emanates from the singularity for a given value of $v_{\rm s}$.
Since $v_{\rm s}$ can take continuous values $v_{\rm s}\in (-\infty,0)$ for the singularity $r=0$ in the untrapped region ($v<0$), it is timelike.
On the other hand, in the trapped region ($f<0$), $r$ is a timelike coordinate and so $\gamma_{\rm out}$ does not emanate from but terminates at the singularity.
Therefore in this case, the singularity at $r=0$ (with $v>0$) is spacelike.

We have seen that the central singularity in the domain of $v<0$ is a timelike naked singularity.
Therefore, in order to construct a model of the gravitational collapse from regular initial data, we attach the Vaidya-AdS spacetime (\ref{vaidya-gr}) for $v>0$ to a locally AdS spacetime for $v<0$ at the matching null hypersurface $v=0$, which is denoted by $\Sigma$. 
This locally AdS spacetime is described by the line element
\begin{equation}
\D s^2=-\frac{r^2}{l^2}\D v^2+2\D v\D r+r^2\D\theta^2, \label{mBTZ}
\end{equation}
which is called as the massless BTZ spacetime written in the ingoing Eddington-Finkelstein coordinates. 

We are going to show that this matching surface is regular, namely there is no massive thin-shell on $\Sigma$.
In the case of a null hypersurface in general relativity, continuity of the induced metric and the transverse curvature of the two matching spacetimes at $\Sigma$ are sufficient for the absence of a massive thin-shell.
(See~\cite{bi1991,Poisson} for the matching condition on a null hypersurface in general relativity.)

On the null hypersurface $\Sigma$ defined by $v=0$, we install coordinates $y^a=({\bar\lambda},\theta^A)$ which are the same on both past and future sides of $\Sigma$.
Here ${\bar \lambda}$ is an arbitrary parameter on the null generators of $\Sigma$ and $\theta^A$ label the generators, where the index $A$ is always $A=1$ in the three-dimensional case.
We identify $-r$ with ${\bar\lambda}$ and set $\theta^A=\theta$ on $\Sigma$ in the spacetime (\ref{metric-assumption}).
The line element on $\Sigma$ is one-dimensional and given by
\begin{align}
\D s_{\Sigma}^2=\sigma_{AB}\D \theta^A \D \theta^B={\bar\lambda}^2\D\theta^2,\label{hab}
\end{align}
where $\sigma_{AB}$ is the induced metric on $\Sigma$.
The parametric equations $x^\mu=x^\mu({\bar\lambda},\theta^A)$ describing $\Sigma$ are $v=0$, $r=-{\bar\lambda}$, and $\theta=\theta$.
Using them, we obtain the tangent vectors of $\Sigma$ defined by $e^\mu_a := \partial x^\mu/\partial y^a$ as
\begin{align}
e^\mu_{\bar \lambda}\frac{\partial}{\partial x^\mu}=-\frac{\partial}{\partial r},\qquad e^\mu_\theta\frac{\partial}{\partial x^\mu}=\frac{\partial}{\partial \theta}
\end{align}
and an auxiliary null vector $N^\mu$ given by
\begin{align}
N^\mu \frac{\partial}{\partial x^\mu}=\frac{\partial}{\partial v}+\frac12f(0,r)\frac{\partial}{\partial r} \label{N-attachment}
\end{align}
completes the basis.
The expression $N_\mu \D x^\mu=-(f(0,r)/2)\D v+\D r$ shows $N_\mu N^\mu =0$, $N_\mu e^\mu_{\bar \lambda}=-1$, and $N_\mu e^\mu_\theta=0$.
Then, the only nonvanishing component of the transverse curvature $C_{ab}:=(\nabla_\nu N_{\mu}) e^\mu_{a} e^\nu_b$ of $\Sigma$ is 
\begin{align}
C_{\theta\theta}=\frac12rf(0,r).\label{trans-C0}
\end{align}
Regular attachment without a massive thin-shell requires continuity of $\sigma_{AB}$ and $C_{ab}$ at $\Sigma$.

Now we attach the Vaidya-AdS spacetime (\ref{vaidya-gr}) with $\mu(v)=\mu_1v$ for $v\ge 0$ to the massless BTZ spacetime, given by $\mu(v)\equiv 0$, for $v\le 0$.
Then, Eqs.~(\ref{hab}) and (\ref{trans-C0}) show that both $\sigma_{AB}$ and $C_{ab}$ are continuous at $v=0$ and hence $\Sigma$ is regular.

We have shown that the Vaidya-AdS spacetime for $v>0$ can be attached to the past massless BTZ spacetime for $v<0$ in a regular manner and the singularity at $v>0$ is spacelike and censored.
Now the remaining problem is whether the point $v=r=0$ is a naked singularity or not.
Since the trapping horizon $v=v_{\rm TH}(r):=r^2/(\mu_1l^2)$ is an increasing function in the $(r,v)$-plane, there may exist future-directed outgoing causal geodesics emanating from the singularity at $v=r=0$.
Such causal geodesics $v=v_{\rm CG}(r)$ satisfies $v_{\rm CG}(r)<v_{\rm TH}(r)$ near $v=r=0$.

In order to clarify the nature of the point $v=r=0$, the contraposition of the following lemma is useful.
(The proof is similar to the four-dimensional case in~\cite{nmg2002}.) 
\begin{lm}
\label{lm:geodesics}
If a future-directed outgoing causal (excluding radial null) geodesic emanates from the singularity, then a future-directed outgoing radial null geodesic emanates from the singularity.
\end{lm}
The contraposition of this lemma asserts that it is sufficient to study outgoing radial null geodesics to prove that the singularity is censored.

Future-directed outgoing radial null geodesics satisfy $\D r/\D v=(r^2/l^2-\mu_1 v)/2$.
This equation is integrated to give 
\begin{align}
r(v)=-\frac{2l^2\eta(c_1\mbox{Ai'}({\eta v})+\mbox{Bi'}({\eta v}))}{c_1\mbox{Ai}({\eta v})+\mbox{Bi}({\eta v})}, \label{Sol-Airy}
\end{align}
where $\eta:=[\mu_1/(4l^2)]^{1/3}$,  $c_1$ is an integration constant, and Ai and Bi are the Airy wave functions.
In Eq.~(\ref{Sol-Airy}), a prime denotes derivative with respect to the argument.
Outgoing radial null geodesics satisfying $r(0)=0$ correspond to $c_1=\sqrt{3}$ and then $r(v)$ behaves near $v=0$ as $r(v)\simeq -\mu_1v^2/4+{\cal O}(v^5)$.
Since this is non-positive near $v=0$, there is no future-directed outgoing radial null geodesic emanating from the singularity at $r=v=0$ and therefore it is censored.

\begin{figure}[htbp]
\begin{center}
%\rotatebox{-90}{
\includegraphics[width=0.3\linewidth]{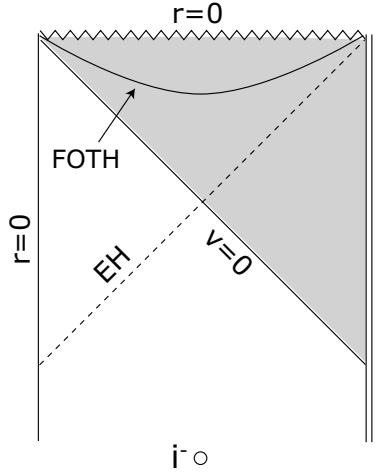}
%\subfigure[]{\includegraphics[width=0.7\linewidth]{Roberts-lambda1.eps}}
%\subfigure[]{\includegraphics[width=0.7\linewidth]{Roberts-lambda2.eps}}
%}
\caption{\label{Fig-VaidyaAdS} The Penrose diagram of the three-dimensional Vaidya-AdS spacetime (shadowed region) attached to the past massless BTZ spacetime at $v=0$.
A dashed line and a solid curve are the event horizon (EH) and a future outer trapping horizon (FOTH), respectively.
A double line is the AdS infinity, while $i^-$ denotes the past timelike infinity.}
\end{center}
\end{figure}

Based on all the information obtained up to now, the Penrose diagram of the resulting spacetime is drawn as Fig.~\ref{Fig-VaidyaAdS}.
It clearly shows that the spacetime represents the black-hole formation in an asymptotically AdS spacetime from regular initial data.
%------------<fig>---------------------------
%--------------<fig>-----------------------

%======================================%
%<<<<<<<<<<<< SECTION I  >>>>>>>>>>>>>>%
%======================================%
\section{Exact solutions}
\label{sec:exact}
In this section, we will present all the possible solutions in the system (\ref{action-1000}) under the metric assumption (\ref{metric-assumption}).
Actually in~\cite{Xu:2014xqa}, Xu solved the field equations under the following different metric assumption:
\begin{equation} \label{ans}
\D s^2=-f(v,r)\D v^2+2\D v \D r+ r^2 h(v)\D  \theta^2,
\end{equation}
where a scalar field is assumed to depend on $v$ and $r$,  i.e.,  $\phi=\phi(v,r)$.
However, we can set $h(v) \equiv 1$ in the metric (\ref{ans}) without loss of generality, as shown below.

In what follows,  the Einstein equations (\ref{Einstein}) are written as ${\cal E}_{\mu\nu}:=G_{\mu \nu}-l^{-2} g_{\mu \nu}- \kappa T_{\mu \nu}=0$.
With the metric (\ref{ans}), an uncoupled equation for $\phi(v,r)$ is provided by ${\cal E}^{v}_{r}=0$ as
\begin{equation}
3 \left(\frac{\partial \phi }{\partial r}\right)^2-\phi  \frac{\partial^2 \phi}{\partial r^2}=0, 
\end{equation}
whose general solution is given by
\begin{equation} \label{phi}
\phi(v,r) =\frac{a(v)}{\sqrt{r+b(v)}},
\end{equation}
where $a(v)$ and $b(v)$ are arbitrary functions.
On the other hand, since the energy-momentum tensor is traceless, we have $ R= -6 l^{-2}$, which leads to
\begin{equation} 
\frac{\partial^2 f}{\partial r^2}+\frac{2}{r}\frac{\partial f }{\partial r}+\frac{2 h'(v)}{r h(v)}=\frac{6}{l^2}.
\end{equation}
The above equation can be completely integrated as
\begin{equation} \label{F}
f(v,r)= \frac{r^2}{l^2}-B(v)-\frac{A(v)}{r}-\frac{r h'(v)}{h(v)},
\end{equation} 
where $A(v)$ and $B(v)$ are arbitrary functions.
Then, we consider the coordinate transformations 
\begin{equation} \label{ct}
\tilde{r} = r h^{1/2}, \qquad  \D \tilde{v} = h^{-1/2}\D v,
\end{equation}
in conjunction with the following redefinitions of the functions
\begin{equation} \label{redefs}
\tilde{B}:= Bh, \quad \tilde{A}:= A h^{3/2}, \quad \tilde{a}:= a h^{1/4}, \quad \tilde{b}:= b h^{1/2}.
\end{equation}
Omitting the tildes, we obtain
\begin{equation} \label{ansh1}
\D s^2=-\left(\frac{{r}^2}{l^2}-{B} -\frac{{A} }{{r}}\right)\D {v}^2+2\D {v} \D {r}+ {r}^2 \D \theta^2
\end{equation}
and
\begin{equation} \label{phih1}
\phi =\frac{{a}}{\sqrt{{r}+{b}}},
\end{equation}
which is a solution in the form of (\ref{ans}) with $h(v) \equiv 1$. Thus, we have shown that all the solutions under the metric assumption (\ref{metric-assumption}) reduce to Eqs.~(\ref{ansh1}) and (\ref{phih1}). Consequently, the on-shell energy-momentum tensor is given by  
\begin{equation} \label{Tonshell}
T^{\mu} \,_{ \nu}=\frac{1}{\kappa}\biggl(G^{\mu} \,_{ \nu}-\frac{1}{l^2}\delta^{\mu} \,_{ \nu}\biggl)=\frac{1}{2 r^3 \kappa } \left(
\begin{array}{ccc}
 A  & 0 & 0 \\
 r A'+r^2 B' & A & 0 \\
 0 & 0 & -2A \\
\end{array}
\right).
\end{equation}
Note that solutions with $A\equiv 0$ and $B'\equiv 0$ yield a vanishing energy-momentum tensor. 
Such a class of solutions will be presented in the next subsection, even in presence of a nonzero scalar field.

For the metric \eqref{ansh1} with the scalar field \eqref{phih1}, the Einstein equations ${\cal E}_{\mu\nu}=0$ are written as
\begin{align}
0=&16 l^2 r^3 (r+b)^3{\cal E}^{v} \,_{ v}=16 l^2 r^3 (r+b)^3{\cal E}^{r} \,_{r} \nonumber \\
=&r^3 \biggl\{-4 \kappa  l^2 a b a'-2 \kappa  a^2 \left(l^2 b'+b^2\right)+16 \kappa \alpha  l^2 a^6+8 l^2 A\biggl\} \nonumber \\ 
&+2 l^2 r^2 b \biggl\{\kappa  a^2 \left(b'-B\right)-2 \kappa  ab a'+12 A\biggl\}+3 l^2 r A b \left(8 b-\kappa  a^2\right)+l^2 A b^2 \left(8 b-\kappa  a^2\right), \label{eqvv}\\
0=&-16 l^2 r^2 (r+b)^2{\cal E}^{r} \,_{v} \nonumber \\
=&-2 r^3 \biggl\{2 l^2 \left(2 B'-3 \kappa  a'^2\right)+a \left(2 \kappa  l^2 a''-2 \kappa  b a'\right)+\kappa  a^2 b'\biggl\} \nonumber \\
&+2 l^2 r^2 \biggl\{b \left(6 \kappa  a'^2-8 B'\right)+2 \kappa  a \left(-b a''-2 a' b'+B a'\right)+\kappa  a^2 b''-4 A'\biggl\}  \nonumber \\
&+l^2 r \biggl\{6 \kappa  a A a'+\kappa  a^2 b B'-8 b \left(2 A'+b B'\right)\biggl\} +l^2 \biggl\{2 \kappa  a A b a'+\kappa  a^2 \left(b A'-A b'\right)-8 b^2 A'\biggl\},\label{eqrv}\\\
0=&8 l^2 r^3 (r+b)^3 {\cal E}^{\theta} \,_{\theta} \nonumber \\
=&r^3 \biggl\{-\kappa  a^2 b^2+\kappa l^2 a^2 \left(B-2 b'\right)+8 \kappa \alpha  l^2 a^6-8 l^2 A\biggl\} \nonumber \\ 
&+3 l^2 r^2 A \left(\kappa  a^2-8 b\right)+3 l^2 r A b \left(\kappa  a^2-8 b\right)+l^2 A b^2 \left(\kappa  a^2-8 b\right). \label{eqpp}\
\end{align}
Thus, the combination ${\cal E}^{\theta}_{~\theta}-{\cal E}^{v}_{~v}=0$ gives 
\begin{eqnarray} \label{pol}
r^2 \biggl\{-2 \kappa  a^2 \left(b'-B\right)+4 \kappa  ab a'-24 A\biggl\}+6 r A \left(\kappa  a^2-8 b\right)+3 A b \left(\kappa a^2-8 b\right)=0.
\end{eqnarray}
This is a quadratic polynomial on $r$ whose coefficients must vanish. 
The order $r^0$ coefficient gives
\begin{equation} \label{3branches}
A b \left(\kappa a^2-8 b\right)=0,
\end{equation}
which implies three classes of solutions with a nontrivial scalar field ($a\neq 0$). 
We will analyse below these three classes from Eq.~\eqref{3branches} and present all the possible solutions.

\subsection{Stealth solutions}
\subsubsection{Class I: $A(v)\equiv 0$} \label{A=0}
First we consider the case where $b\neq 0$ and $\kappa  a^2 \neq 8 b$ are satisfied in addition to $A\equiv 0$. 
The linear ($r^1$) term in \eqref{eqrv} reduces to $l^2 r b \left(\kappa  a^2  -8 b \right)B'$,  so that  $B'=0$ and hence $B(v) =M_0$, where $M_0$ is a constant.
Thus, the metric function $f$ in this class is given by 
\begin{equation} \label{F1}
f(r)=\frac{r^2}{l^2}-M_0.
\end{equation}
The energy-momentum \eqref{Tonshell} vanishes because of $A\equiv 0$ and $B' = 0$, and consequently the Riemann  tensor takes the form $R^{\mu \nu}_{~~~\lambda \rho}=-l^{-2}(\delta^\mu_\lambda\delta^\nu_\rho-\delta^\mu_\rho\delta^\nu_\lambda)$ since we are dealing with a three-dimensional Einstein spacetime. 
Therefore, this class yields a \textit{stealth} scalar field\footnote{We adopt the same name for this configuration as it was coined in \cite{AyonBeato:2004ig}. Stealth solutions are not exclusives of three-dimensional gravity but also they exist in arbitrary dimensions \cite{AyonBeato:2005tu}.}, namely, a nontrivial field with vanishing energy-momentum tensor. 

The quadratic ($r^2$) term in \eqref{pol} gives the following differential equation:
\begin{equation} 
b'=\frac{2 b a'}{a} +M_0. \label{b1}  
\end{equation}
Replacing $b'$ and $b''$ in Eqs.~\eqref{eqvv}--\eqref{eqpp} by using Eq.~\eqref{b1}, we obtain 
\begin{equation} \label{hh1simple}
(a^{-2})''-M_0 l^{-2} a^{-2}=0
\end{equation}
with a constraint equation
\begin{eqnarray} 
\frac{a'}{a}&=&\frac{2 \alpha  a^4}{b}-\frac{M_0}{4 b}-\frac{b}{4 l^2}.\label{h1}
\end{eqnarray}

In the case of $M_0 > 0$, the general solution of Eqs.~(\ref{b1})--(\ref{h1}) is
\begin{align}
a(v)=&\pm\left(a_0 \cosh \left[\frac{\sqrt{M_0} (v-v_0)}{l}\right]\right)^{-1/2}, \\ 
b(v)=&l \sqrt{M_0} \biggl(\cosh\left[\frac{\sqrt{M_0} (v-v_0)}{l}\right]\biggl)^{-1} \left(\frac{b_0}{a_0}+ \sinh \left[\frac{\sqrt{M_0} (v-v_0)}{l}\right]\right),
\end{align}
where the integration constants $a_0$ and $b_0$ satisfy $(a_0^2+b_0^2)M_0=8 \alpha$ and  $v_0$ is an arbitrary constant. 
These $a$ and $b$ give the following form of the scalar field:
\begin{align}
\phi(v,r)=\pm\biggl[a_0r\cosh\left[\frac{\sqrt{M_0} (v-v_0)}{l}\right]+l \sqrt{M_0} \left(b_0+ a_0\sinh \left[\frac{\sqrt{M_0} (v-v_0)}{l}\right]\right)\biggl]^{-1/2}. \label{stealth-scalar1}
\end{align}
This is a stealth configuration overflying a static BTZ black hole (\ref{F1}) with its mass $M_0(>0)$, which was found in \cite{AyonBeato:2004ig} using the following time coordinate $t$: 
\begin{equation} \label{tc}
t =v-\int\frac{\D r}{f(r)}=v+\frac{l}{\sqrt{M_0}}\mbox{arctanh}\left(\frac{r}{l \sqrt{M_0}}\right).
\end{equation}
Note that this stealth solution for $M_0>0$ is supported by a non-vanishing potential ($\alpha \neq 0$). 
In contrast, the stealth solutions for $M_0 \leq 0$ presented below are possible even in the absence of a potential ($\alpha=0$).

In the case of $M_0 < 0$, the general solution of Eqs.~(\ref{b1})--(\ref{h1}) is
\begin{align}
a(v)=&\pm\left(a_0 \cos \left[\frac{\sqrt{-M_0} (v-v_0)}{l}\right]\right)^{-1/2}, \\ 
b(v)=&l \sqrt{-M_0} \biggl(\cos\left[\frac{\sqrt{-M_0} (v-v_0)}{l}\right]\biggl)^{-1} \left(\frac{b_0}{a_0}- \sin \left[\frac{\sqrt{-M_0} (v-v_0)}{l}\right]\right),
\end{align}
where integration constants $a_0$ and $b_0$ satisfy $(a_0^2-b_0^2)M_0=8 \alpha$. Then, the scalar field becomes
\begin{align}
\phi(v,r)=\pm\biggl[a_0r\cos\left[\frac{\sqrt{-M_0} (v-v_0)}{l}\right]+l \sqrt{-M_0} \left(b_0- a_0\sin \left[\frac{\sqrt{-M_0} (v-v_0)}{l}\right]\right)\biggl]^{-1/2}. \label{stealth-scalar1-2}
\end{align}

In the case of $M_0 = 0$, the general solution of Eqs.~(\ref{b1})--(\ref{h1}) is
\begin{equation} \label{am0}
a(v)= \pm\frac{1}{\sqrt{a_0 v-v_0}},\qquad b(v)= \frac{b_0}{a_0 v-v_0},
 \end{equation}
where $a_0, b_0$  and $v_0$ are integrations constants satisfying 
\begin{equation} \label{cc1}
b_0^2-2a_0  b_0 l^2 = 8 \alpha l^2.
\end{equation}
Then, the scalar field takes the form of
\begin{align}
\phi(v,r)=\pm\frac{1}{\sqrt{r (a_0 v-v_0)+b_0}}.\label{stealth-scalar1-3}
\end{align}
Different from the case of $M_0\ne 0$, this solution (\ref{stealth-scalar1-3}) admits a static stealth configuration, given by $a_0=0$:
\begin{align}
\phi(r)=\pm\frac{1}{\sqrt{- v_0 r\pm\sqrt{8 \alpha l^2}}},\label{stealth-scalar1-4}
\end{align}
where two $\pm$ are independent.
This static stealth solution for $M_0=0$ \cite{Gegenberg:2003jr} is compatible only with a non-negative coupling constant $\alpha \ge 0$.

Next, we consider the case where $A\equiv 0$ and $b\equiv \kappa  a^2/8\neq 0$ are satisfied.
Then Eqs.~\eqref{eqvv}--\eqref{eqpp} reduce to
\begin{align}
0=&  \left(4 a'-\beta  a^3\right)r+a B,  \label{41}\\
0=& 2 \left(\kappa  a a''-3 \kappa  a'^2+2 B'\right)r +\kappa a ( a B'-2 a' B ), \label{42}\\
0=&\kappa  a (4  a'-\beta   a^3) -8 B\label{43},
\end{align}
where $\beta$ is defined by Eq.~(\ref{def-beta}).
Eq.~\eqref{41} implies $B = 0$, so that we have only two independent equations:
\begin{align}
4 a'&=\beta  a^3,  \label{51}\\
a a''&=3  a'^2. \label{52}
\end{align}
If $\beta=0$, then the static scalar field \eqref{stealth-scalar1-4} is obtained.
If $\beta\neq 0$, we obtain 
\begin{equation} 
a(v)=\pm\sqrt{-\frac{2}{\beta v+v_0}},
\end{equation}
where $v_0$ is an integration constant. 
Thus, the metric function $f$ and scalar field $\phi$ are given by 
\begin{equation} \label{53}
f(r)=\frac{r^2}{l^2}, \qquad \phi(v,r)=\pm\sqrt{\frac{8}{\kappa- 4(\beta v+v_0)r}},
\end{equation}
which work for any value of $\beta$.

There remains a case where both $A\equiv 0$ and $b\equiv 0$ are satisfied.
Because $b\equiv 0$ implies $A = 0$, this case will be treated in the following subsection.

\subsubsection{Class II: $b(v)\equiv 0$}
In the case of $b(v)\equiv 0$, Eq.~\eqref{pol} directly implies that both $A(v) =0$ and $B(v)= 0$ hold, and the remaining equations are consistent only in the absence of a self-interaction potential, namely with $\alpha=0$. 
Then, the system reduces to the following single equation for $a$:
\begin{equation}
a ''-\frac{3 a'^2}{a}=0, \label{hb}
\end{equation}
whose solution is given by
\begin{equation}
a(v)= \frac{a_0}{\sqrt{v-v_0}},
\end{equation}
where $a_0$ and $v_0$ are constants.
Thus, the metric function $f$ and scalar field $\phi$ become
\begin{eqnarray}
f(r) =\frac{r^2}{l^2},\qquad \phi(v,r) = \frac{a_0}{\sqrt{r(v-v_0)}}. \label{pb}
\end{eqnarray}

In fact, this solution (\ref{pb}) can be obtained from the massless BTZ black hole with a stealth scalar field given in~\cite{AyonBeato:2004ig} by the coordinate transformation $t=v+l^2/r$.  
Moreover, this configuration \eqref{pb} coincides with \eqref{stealth-scalar1-3} if $b_0=0$, which consistently implies $\alpha=0$ by virtue of the constraint \eqref{cc1}.

\subsection{Non-stealth solutions in the class of $b(v)\equiv \kappa a(v)^2/8$}
In the last class, where $b(v)\equiv \kappa a(v)^2/8$ holds, we obtain from \eqref{pol} the following relation
\begin{equation}
A(v)=\frac{1}{12}\kappa  a(v)^2 B(v),
\end{equation}
and then Eqs.~\eqref{eqvv}--\eqref{eqpp} reduce to 
\begin{align}
0=& 24  \left(\kappa  a a''-3 \kappa  a'^2+2 B'\right) r^2+\left(16 \kappa  a r+\kappa ^2 a^3 \right) \left(a B'-B a' \right), \label{erv31}  \\
0=& -12 \kappa   aa'+3 \beta  \kappa  a^4+8 B,  \label{evv31}
\end{align}
where $\beta$ is defined by Eq.~(\ref{def-beta}).
The first of the above equations yields
\begin{equation}
B(v)=B_0 a(v),
\end{equation}
where $B_0$ is an integration constant. 
Then, Eqs.~\eqref{erv31}--\eqref{evv31} become 
\begin{align}
\kappa  a a''=& 3 \kappa  a'^2-2 B_0 a', \label{h30} \\
12 \kappa   a'=&3 \beta  \kappa  a^3+8 B_0. \label{h3}
\end{align}
Eq.~\eqref{h3} is the master equation for $a(v)$ because any solution of Eq.~\eqref{h3} solves Eq.~\eqref{h30}. 

Thus,  we have shown that the solution in this class is given by
\begin{eqnarray} \label{sol3f}
f(v,r)&=& \frac{r^2}{l^2}-B_0 a(v)-\frac{B_0 \kappa  a(v)^3}{12 r },\\
\phi(v,r) &=& \frac{a(v)}{\sqrt{r+\kappa a(v)^2/8}},\label{sol3p}
\end{eqnarray}
where $a(v)$ is governed by Eq.~(\ref{h3}).
Actually, this solution with $B_0\ne 0$ is the only non-stealth configuration in the present system under the metric assumption \eqref{ansh1}.
The stealth configuration given by Eqs.~(\ref{sol3f}) and (\ref{sol3p}) with $B_0=0$ is identical to the solution \eqref{53} obtained in Sec.~\ref{A=0}.
Now we solve the master equation~(\ref{h3}) for $a(v)$ with $B_0\ne 0$.

\subsubsection{Static case: Henneaux-Mart\'{\i}nez-Troncoso-Zanelli (HMTZ) solution}
First let us consider the static solution of the master equation \eqref{h3}, namely $a=a_0$.
The constant $a_0$ is given by 
\begin{equation} 
a_0=-\epsilon\biggl|\frac{8B_0 }{3 \kappa \beta}\biggl|^{1/3},\label{a0-def}  
\end{equation}
where $\epsilon$ is the sign of $8B_0/(3 \kappa \beta)$.
Using the relation 
\begin{equation} 
B_0=-\frac{3}{8} \kappa \beta a_0^3,\label{B0-beta}  
\end{equation}
we can write the solution as
\begin{equation} \label{statEF}
f(r)= \frac{r^2}{l^2}-\biggl(1-\frac{512 \alpha l^2}{\kappa^2}\biggl) \left(\frac{3 \kappa^2 a_0^4}{64 l^2}+\frac{\kappa^3 a_0^6}{256l^2 r} \right), \quad \phi(r) = \frac{a_0}{\sqrt{r+\kappa a_0^2/8}},
\end{equation}
where $a_0$ is treated as an integration constant.
This solution, exhibited here in the ingoing Eddington-Finkelstein  coordinates, is identical to the one previously obtained in~\cite{Martinez:1996gn} (for $\alpha=0$) and~\cite{Henneaux:2002wm} (for $\alpha\ne 0$), which is written in the standard diagonal coordinates as
\begin{equation}
\D s^2=-F(r)\D t^2+\frac{\D r^2}{F(r)}+r^2\D \theta^2, \quad \phi=\phi(r),\label{HMTZ}
\end{equation}
where 
\begin{equation} \label{statold}
F(r)=\frac{r^2}{l^2}- \left(1-\frac{512\alpha  l^2}{\kappa^2}\right) \left(\frac{3C^2}{l^2}+\frac{2C^3}{l^2 r} \right), \quad \phi(r)=\sqrt{\frac{8C}{\kappa(r+C)}}.
\end{equation}
The relation between the integration constants $C$ and $a_0$ is $C=\kappa a_{0}^2 /8$. 
This static configuration \eqref{statold} represents an asymptotically AdS hairy black hole dressed with a regular scalar field provided $C>0$ and 
\begin{equation}
1-\frac{512\alpha  l^2}{\kappa^2} = -\frac{8 l^2}{\kappa} \beta >0. \label{BH-condition-static}
\end{equation}

\subsubsection{Dynamical case: Generalization of the Xu's solution}
Now let us see non-static solutions of the master equation~\eqref{h3}.
In the case of $\beta=0$, namely, if there is a fine-tuning between the cosmological constant and the potential parameter such that 
\begin{eqnarray}
512\alpha l^2 =\kappa^2,
\end{eqnarray}
the general non-static solution of Eq.~(\ref{h3}) is 
\begin{eqnarray}
a(v)=\frac{2B_0 }{3 \kappa }(v-v_0),\label{param-beta=0}
\end{eqnarray}
where $v_0$ is a constant.

On the other hand, the general non-static solution of \eqref{h3} for $\beta\ne 0$ is 
\begin{equation} \label{param}
\frac12\ln \left(\frac{(a-a_0)^2}{a^2+a_0 a+a_0^2}\right)-\sqrt{3} \arctan\left(\frac{2 a+a_0}{\sqrt{3} a_0}\right)=\frac{3 a_0^2 \beta }{4} (v-v_0),
\end{equation}
where $v_0$ is an integration constant and $a_0$, defined by Eq.~(\ref{a0-def}),  is the value of $a$ for the static solution. A detailed analysis of this non-stealth solution is presented in the next section.

Now let us show the relation between our solution (\ref{param}) and the solution obtained by Xu~\cite{Xu:2014xqa}.
Xu's metric is described (in a different notation) by 
\begin{align}
\D s^2=&-H(u,{\bar r})\D u^2+2\D u\D {\bar r}+{\bar r}^2\tanh^{2/3}\biggl(\frac{12{\bar \alpha} u}{q}\biggl)\D \theta^2,\label{Xu-metric}\\
H(u,{\bar r})=&\frac{{\bar r}^2}{l^2}-\frac{12{\bar\alpha}}{q^2}-\frac{{\bar\alpha}}{q^3{\bar r}}\tanh\biggl(\frac{12{\bar \alpha} u}{q}\biggl)+\frac{8{\bar\alpha}\{\tanh^{2}(12{\bar \alpha} u/q)-1\}}{q\tanh(12{\bar \alpha} u/q)}{\bar r} \label{hXu}
\end{align}
and the scalar field is 
\begin{equation} \label{sXu}
\phi(u,{\bar r})=\sqrt{\frac{8 \tanh(12 \bar{\alpha} u/q)}{ 8 q \bar{r} +\tanh (12 \bar{\alpha} u/q)}},
\end{equation}  
where he adopted the units such that $\kappa=1$.
With $\kappa=1$, the constant ${\bar \alpha}$ (which is written as $\alpha$ in his paper) is related to our $\alpha$ and $l$ as
\begin{equation}
{\bar \alpha}=\frac{1-512\alpha l^2}{256l^2}=-\frac{\beta}{32}. \label{alpha-bar}
\end{equation}

In his paper, Xu claims that, since the spacetime is AdS at $u=0$ and there is no singularity in the domain of $u\in[0,\infty)$, the solution represents gravitational collapse from AdS converging to a static black hole in far future $u\to+\infty$. However, this claims is invalid because the metric (\ref{Xu-metric}) has a coordinate singularity at $u=0$ as it can be seen in Eq.~\eqref{hXu}. Therefore, 
the correct domain of $u$ is given in the metric (\ref{Xu-metric}) is ${\bar \alpha}u/q\in(0,\infty)$.

As shown in Eq.~(\ref{ct}), the metric (\ref{Xu-metric}) is transformed from the coordinate system $(u,{\bar r},\theta)$ into our coordinate system $(v,r,\theta)$ by the following transformations:
\begin{equation} 
r={\bar r} \tanh^{1/3}\biggl(\frac{12{\bar \alpha} u}{q}\biggl),\qquad v=\int \tanh^{-1/3}\biggl(\frac{12{\bar \alpha} u}{q}\biggl)\D u,
\end{equation}
from which we obtain 
\begin{align}  \label{vu}
\frac12\ln\biggl(\frac{(U^{2/3}-1)^2}{U^{4/3}+ U^{2/3}+1}\biggl) -\sqrt{3}\arctan\biggl(\frac{2U^{2/3}+1}{\sqrt{3}}\biggl)=-\frac{24{\bar \alpha}}{q}(v-v_1),
\end{align}
where $v_1$ is an integration constant and $U:=\tanh (12{\bar \alpha} u/q) \in (0,1)$.  
This shows that ${\bar \alpha}u/q\to +\infty$ corresponds to ${\bar \alpha}v/q\to +\infty$ and therefore $u\to +\infty$ in Xu's paper is our $v\to +\infty$ limit.

Comparing Eq.~(\ref{sol3f}) with Eq.~(\ref{hXu}) and also Eq.~(\ref{param}) with Eq.~(\ref{vu}), we identify $q=1/a_0^2$, $v_1=v_0$, and 
\begin{align} \label{avu}
a(v(u))=&\frac{12{\bar\alpha}}{q^2 B_0}\tanh^{2/3}\biggl(\frac{12{\bar \alpha} u}{q}\biggl)=a_0U^{2/3},
\end{align}   
where we used Eq.~(\ref{alpha-bar}).
The function $a$ in Xu's coordinates \eqref{avu} is restricted to have a single sign once $\bar\alpha$ and $B_0$ are fixed because of $\tanh^{2/3}(12{\bar \alpha} u/q) >0$. 
In contrast, the function $a$ in our coordinates is not limited to have a definite sign, as shown is the following section. 
Therefore, our solution (\ref{param}) is an analytic extension of Xu's solution beyond $u=0$.

It is also noted that the limit ${\bar \alpha}u/q\to 0$ in Eq.~\eqref{vu} corresponds to the following finite value of $v$:
\begin{align} 
v=v_1+\frac{\sqrt{3}\pi q}{144{\bar \alpha}}=v_0-\frac{2\sqrt{3}\pi}{9\beta a_0^2}=:v_{\rm min}. \label{v-minimum}
\end{align}
Thus, Xu's coordinates (\ref{Xu-metric}) do not cover the domain of $v\le v_{\rm min}$ in our coordinate system.

%======================================%
%<<<<<<<<<<<< SECTION I  >>>>>>>>>>>>>>%
%======================================%
\section{Gravitational collapse of a conformally coupled scalar field}
\label{sec:main}

In this section, we present a physical model of the gravitational collapse of a conformally coupled scalar field in an asymptotically AdS (in the form of the massless BTZ) spacetime, constructed from the following non-stealth solution obtained in the previous section:
\begin{align}
\D s^2=&-f(v,r)\D v^2+2\D v\D r+r^2\D\theta^2,\label{sol-attach} \\
f(v,r)=&\frac{r^2}{l^2}-B_0 a(v)-\frac{B_0 \kappa  a(v)^3}{12r}, \label{def-f0}\\
\phi(v,r)=&\frac{a(v)}{\sqrt{r+\kappa a(v)^2/8}}.\label{def-phi0}
\end{align}
The function $a(v)$ is given by Eq.~(\ref{param-beta=0}) for $\beta=0$ and by Eq.~(\ref{param}) for $\beta\ne 0$. 
Without loss of generality, by a shift transformation of $v$, we set $a(0)=0$. This  is equivalent to choose the integration constant $v_0$ in the expression of $a(v)$  as Eq.~(\ref{v0-normal}) for $\beta\ne 0$ and to set $v_0=0$ in the case of $\beta=0$. Since $B_0=0$ gives the massless BTZ spacetime (with or without a stealth scalar field), we assume $B_0\ne 0$ in this section.

\subsection{Properties of the dynamical non-stealth solution}

\subsubsection{Asymptotic behavior for $r\to \infty$}
The AdS spacetime is the maximally symmetric spacetime with negative curvature and the Riemann tensor is given by 
\begin{align}
R^{\mu\nu}_{~~~\rho\sigma}=-\frac{1}{l^2}(\delta^\mu_\rho\delta^\nu_\sigma-\delta^\mu_\sigma\delta^\nu_\rho).
\end{align}
In our solution (\ref{sol-attach}), the nonzero components of the Riemann tensor behave near $r\to \infty$ on a null hypersurface with constant $v$ as
\begin{align}
R^{vr}_{~~~vr}=&\frac{B_0\kappa l^2a^3-12r^3}{12l^2r^3}\to -\frac{1}{l^2}+{\cal O}(r^{-3}),\\
R^{v\theta}_{~~~v\theta}=&R^{r\theta}_{~~~r\theta}=-\frac{B_0\kappa l^2a^3+24r^3}{24l^2r^3}\to -\frac{1}{l^2}+{\cal O}(r^{-3}),\\
R^{r\theta}_{~~~v\theta}=&\frac{B_0a'(\kappa a^2+4r)}{8r^2}\to 0+{\cal O}(r^{-1}).
\end{align}
Therefore, the spacetime is asymptotically (at least locally) AdS for $r\to \infty$ with constant $v$, which is the past null infinity.

\subsubsection{Energy conditions}
\label{sec:energy} 
With the metric function (\ref{def-f0}), namely
\begin{align}
f(v,r)=&\frac{r^2}{l^2}-B(v)-\frac{A(v)}{r}
\end{align}
with
\begin{align}
A(v)=\frac{B_0 \kappa  a(v)^3}{12},\qquad B(v)=B_0a(v),
\end{align}
the energy-momentum tensor for the conformally coupled scalar field is given by 
\begin{equation} 
\label{Tmunu}
T_{\mu\nu}=\frac{1}{\kappa}\biggl(G_{\mu\nu}-\frac{1}{l^2}g_{\mu\nu}\biggl)=\frac{1}{2 \kappa r^3 } \left(
\begin{array}{ccc}
 \Omega  & A & 0 \\
 A & 0 & 0 \\
 0 & 0 & -2r^2A \\
\end{array}
\right),
\end{equation}
where
\begin{equation}
\Omega:=rA'+r^2B'-A\biggl(\frac{r^2}{l^2}-B-\frac{A}{r}\biggl).
\end{equation}

The components of $T_{\mu\nu}$ in the orthonormal frame (\ref{normal-bases}) are computed to give
\begin{equation} 
T^{(a)(b)}=\eta^{(a)(c)}\eta^{(b)(d)}T_{\mu\nu}E^\mu_{(c)}E^\nu_{(d)}=\left(
\begin{array}{ccc}
 \sigma+\nu  & \nu & 0 \\
 \nu & -\sigma+\nu & 0 \\
 0 & 0 & p \\
\end{array}
\right), \label{T(a)(b)}
\end{equation}
where 
\begin{align}
\sigma:=-\frac{A}{2\kappa r^3},\qquad \nu:=\frac{A'+rB'}{4\kappa r^2},\qquad p:=-\frac{A}{\kappa r^3}.
\end{align}
This is the three-dimensional version of the type II energy-momentum tensor in the Hawking-Ellis classification~\cite{Hawking:1973uf} and reduces to the type I energy-momentum tensor if $\nu\equiv 0$ holds, which only occur for the non-stealth static solutions presented in the previous section.
According to the result in~\cite{energyconditions}, the standard energy conditions for the type II energy-momentum tensor (\ref{T(a)(b)}) in three dimensions are equivalent to
\begin{itemize}
\item Null energy condition (NEC): $\nu \ge 0$ and $\sigma+p\ge 0$,
\item Weak energy condition (WEC): $\nu \ge 0$, $\sigma+p\ge 0$, and $\sigma\ge 0$,
\item Strong energy condition (SEC): $\nu \ge 0$ and $p\ge |\sigma|$,
\item Dominant energy condition (DEC): $\nu \ge 0$ and $\sigma\ge |p|$,
\end{itemize}
which include the cases of type I for $\nu\equiv 0$.
(See also~\cite{mmv2017} for the energy conditions in four dimensions.)
In our solution (\ref{sol-attach}), DEC is violated everywhere.
On the other hand, NEC, WEC, and SEC give the same inequalities $A\le 0$ and $A'+rB' \ge 0$, or equivalently $B_0a\le 0$ and $B_0 a'  \ge 0$.
This implies that $aa'>0$ is sufficient to show the violation of NEC, WEC, and SEC independent of the sign of $B_0$.

\subsubsection{Trapping horizon}
\label{Sec:THsub}
Now let us study the properties of the trapping horizon in our solution. In the non-stealth solution \eqref{sol-attach},
the location of the trapping horizon $r=r_{\rm h}(v)$ is determined by $f(r_{\rm h})=0$, namely
\begin{eqnarray}
\frac{r_{\rm h}^2}{l^2}-B_0 a(v)-\frac{B_0 \kappa  a(v)^3}{12r_{\rm h}}=0.\label{def-h}
\end{eqnarray}
This equation shows that, on a null hypersurface with constant $v$ satisfying $B_0a<0$, there is no trapping horizon and $r=0$ is located in the untrapped region.
On the other hand, if $B_0a>0$ is satisfied, Eq.~(\ref{def-h}) allows one real root and hence there is a single trapping horizon and $r=0$ is located in the trapped region.

Equation~(\ref{def-h}) shows the asymptotic behaviors of the trapping horizon.
Near $r_{\rm h}=0$, the relation between $a(v)$ and $r_{\rm h}$ is given by 
\begin{eqnarray}
a(v)\simeq \frac{r_{\rm h}^2}{B_0l^2}. \label{a-r-TH}
\end{eqnarray}
On the other hand, $r_{\rm h}\to \infty$ is realized if and only if $B_0a\to +\infty$.
For $r_{\rm h}\to \infty$, the trapping horizon behaves as 
\begin{eqnarray}
r_{\rm h}\simeq \mbox{sign}(B_0)\biggl|\frac{B_0 \kappa l^2}{12}\biggl|^{1/3}a \to \infty. \label{rh-infity}
\end{eqnarray}

It is shown that the trapping horizon in our solution is future and outer.
The trapping horizon is future and outer if $\partial f/\partial r|_{r=r_{\rm h}}>0$ holds.
This condition is written as
\begin{eqnarray}
B_0 a(v)\biggl(1+\frac{\kappa  a(v)^2}{8r_{\rm h}}\biggl)>0.
\end{eqnarray}
Because the existence of a trapping horizon requires $B_0a>0$ in our solution, it is a future outer trapping horizon.
Nevertheless, the trapping horizon in our solution may show pathological behaviors in the region where $aa'<0$ holds, as seen below.

The line element along the orbit of the trapping horizon $r=r_{\rm h}(v)$ is 
\begin{equation} \label{spaceds2}
\D s^2|_{r=r_{\rm h}(v)}=2\frac{\D r_{\rm h}}{\D v}\D v^2+r_{\rm h}^2\D\theta^2,
\end{equation}
where we used $f(r_{\rm h})=0$. 
Thus, the trapping horizon is spacelike if and only if its area is increasing, namely
\begin{equation} \label{spaceds2f}
\frac{\D r_{\rm h}}{\D v} > 0.
\end{equation}
Actually, from Eq.~(\ref{def-h}), we obtain the following relation:
\begin{eqnarray}
\frac{a}{a'r_{\rm h}}\frac{\D r_{\rm h}}{\D v}= \frac{4r_{\rm h}+\kappa  a^2}{8r_{\rm h}+\kappa  a^2}.
\end{eqnarray}
Because the right-hand side is positive, the above expression and Eq.~(\ref{spaceds2}) show that the area of the trapping horizon is increasing (decreasing) and the trapping horizon is spacelike (timelike) for $aa'>(<)0$.
This means that a future outer trapping horizon in the region of $aa'>0$ is a one-way membrane being matched to the concept of a black hole as a region of no escape.

In contrast, the properties of a future outer trapping horizon in the region of $aa'<0$ are quite different.
In spite that it is an inner boundary of untrapped surfaces, it is not a one-way membrane. 
Indeed, a light ray emanating from a point on a future outer trapping horizon in the region of $aa'<0$ can propagate into both its inside and outside because it is timelike. 
In general relativity, it is shown that, in this class of symmetric spacetimes, an outer trapping horizon is non-timelike under NEC~\cite{hayward1994,nm2008}. 
In our solution, according to the result in Sec.~\ref{sec:energy}, the pathological behavior of the future outer trapping horizon in the region of $aa'<0$ stems from the violation of NEC for the conformally coupled scalar field.

However, it should be emphasized that NEC ensures non-timelikeness of an outer trapping horizon but the opposite is not always true.
Actually, a future outer trapping horizon may be spacelike in the region where NEC is violated.
This is because non-timelikeness of an outer trapping horizon is equivalent to $T_{vv}\ge 0$ in the double-null coordinates $(u,v,\theta)$. (See Lemma 1 and Proposition 10 in~\cite{nm2008}.)
In our coordinate system, this inequality is equivalent to $T_{\mu\nu}k^\mu k^\nu\ge 0$, where $k^\mu$ is the tangent vector of the future-directed outgoing radial null geodesic (\ref{tangent-geodesics}).
Combined with Eq.~(\ref{Tmunu}), we obtain
\begin{equation}
T_{\mu\nu}k^\mu k^\nu=\frac{1}{2\kappa r^2}(A'+rB')
\end{equation}
and hence $A'+rB'\ge 0$ is a necessarily and sufficient condition in order for an outer trapping horizon to be non-timelike.
Since NEC is equivalent to $A\le 0$ and $A'+rB'\ge 0$ as shown in Sec.~\ref{sec:energy}, NEC is (actually all the standard energy conditions are) violated in the region where $A>0$ and $A'+rB'\ge 0$ hold but a future outer trapping horizon is spacelike there.

\subsubsection{Curvature singularities}
The Kretschmann invariant \eqref{RandK} for our solution is calculated to give
\begin{equation} 
K=\frac{12}{l^4}+ \frac{B_0^2 \kappa ^2 a^6}{24 r^6 },\label{K-sol}
\end{equation}
which shows that there is a curvature singularity at the physical center $r=0$ with $a(v)\ne 0$.
The behavior of $K$ for $r=0$ with $a(v)\to 0$ depends on the curve which terminates there.
In addition, there is another curvature singularity when $a(v)\to\pm \infty$ is realized with some value of $v$.

We compute the functions in Eq.~(\ref{L-function}) as
\begin{align}
w=&2r\biggl(\frac{r^3}{l^2}-B_0 ar-\frac{B_0 \kappa  a^3}{12}\biggl)^{-1},\label{h-condition1}\\
w_{,v}=&\frac12B_0r\biggl(\beta a^3+\frac{8B_0}{3\kappa}\biggl)\biggl(r+\frac{\kappa  a^2}{4}\biggl)\biggl(\frac{r^3}{l^2}-B_0 ar-\frac{B_0 \kappa  a^3}{12}\biggl)^{-2},\label{h-condition2}
\end{align}
where we used Eq.~(\ref{h3}).
Both $w$ and $w_{,v}$ are continuous and finite at $r=0$ with $a(v)\ne 0$.
Hence, the central singularity at $r=0$ in the trapped region ($f<0$) is censored and spacelike, while it is naked and timelike if it is located in the untrapped region ($f>0$).
%-------------- TABLE ---------------%
\begin{table}[htb]
\begin{center}
\caption{\label{table:numberTH} Nakedness and signature of the central singularity at $r=0$. The nature of the singularity at $r=0$ with $a=0$ is studied in Sec.~\ref{Sec:r=v=0}.}
\begin{tabular}{|c||c|c|c|c|}
\hline \hline
 & $a<0$ & $a=0$ & $a>0$     \\\hline
$B_0>0$ & Naked (timelike) & $-$ & Censored (spacelike)   \\ \hline
$B_0<0$ & Censored (spacelike) & $-$ & Naked (timelike)  \\ 
\hline \hline
\end{tabular}
\end{center}
\end{table} 
%------------------------------------%

The properties of the central singularity at $r=0$ are summarized in Table~\ref{table:numberTH}.
It is a subtle problem whether the central singularity with $a(v)\to 0$ is naked or not, which will be studied later.

\subsubsection{Behavior of the function $a(v)$}
In our solution, the behavior of the function $a(v)$ is quite nontrivial for $\beta\ne 0$.
Here we present again the master equation \eqref{h3} for $a(v)$:
\begin{equation}
12 \kappa   a'=3 \beta  \kappa  a^3+8 B_0. \label{master-a}
\end{equation}
We are interested in the solution with $B_0\ne 0$.
%-------------- TABLE ---------------%
\begin{table}[htb]
\begin{center}
\caption{\label{table:domain-v3} The behavior of $a(v)$ for $\beta=0$ with $B_0\ne 0$. The symbols $\nearrow$ and $\searrow$ mean that $a$ is increasing and decreasing, respectively.
Also, T and U mean that the central singularity at $r=0$ is in the trapped region and untrapped region, respectively.}
\vskip 3mm
\begin{tabular}{|c||c|c|c|c|c|c|}
\hline \hline
 $v$ & $-\infty$ & $\cdots$ & $0$ & $\cdots$ & $+\infty$ \\\hline
$a(v)~(B_0>0)$ & $-\infty$ & $\nearrow$ (U) & $0$ & $\nearrow$ (T)  & $+\infty$ \\ \hline
$a(v)~(B_0<0)$ & $+\infty$ & $\searrow$ (U) & $0$ & $\searrow$ (T) & $-\infty$ \\ 
\hline \hline
\end{tabular}
\end{center}
\end{table} 
%------------------------------------%

\begin{figure}[htbp]
\begin{center}
%\rotatebox{-90}{
\includegraphics[width=0.7\linewidth]{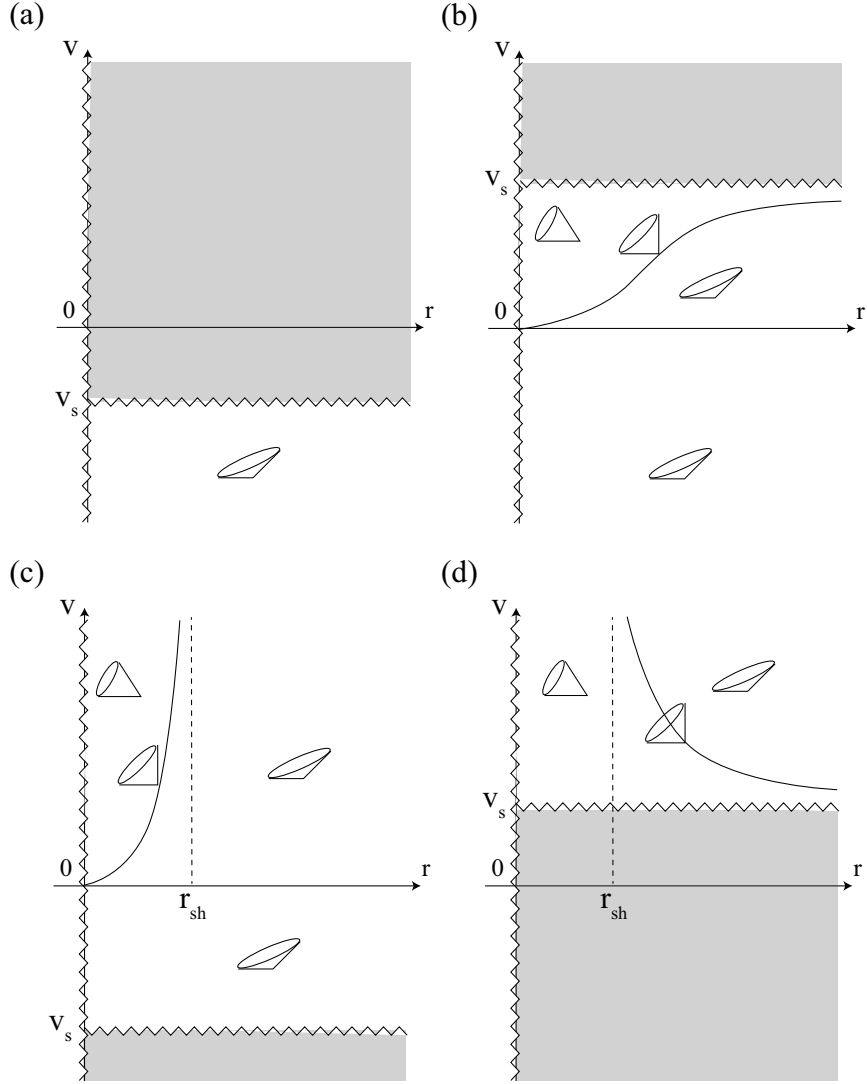}
%\subfigure[]{\includegraphics[width=0.7\linewidth]{Roberts-lambda1.eps}}
%\subfigure[]{\includegraphics[width=0.7\linewidth]{Roberts-lambda2.eps}}
%}
\caption{\label{Fig-vr-plane-all} $(r,v)$-planes for the solution (\ref{solution-a1}) with $\beta\ne 0$, where $v_0$ is given by Eq.~(\ref{v0-normal}).
$v_{\rm s}$ represents $v_+$ or $v_-$ and the corresponding $(r,v)$-planes depending on the parameters are shown in Table~\ref{table:vr}.
The metric is not real in the shadowed regions.
A zigzag line and a solid curve represent a curvature singularity and a future outer trapping horizon, respectively.
The spacetime is asymptotically static for $v\to +\infty $ in (c) and (d) and for $v\to-\infty$ in (a) and (b).
$r=r_{\rm sh}$ is the location of the horizon in the static solution. }
\end{center}
\end{figure}

As shown in Eq.~(\ref{param-beta=0}), the general solution for $\beta=0$ is quite simple:
\begin{eqnarray}
a(v)=\frac{2B_0}{3 \kappa}(v-{v}_0), \label{solution-a0}
\end{eqnarray}
where ${v}_0$ is a constant and we set ${v}_0=0$ without loss of generality by the shift transformation of $v$. 
Then, the behavior of $a(v)$ for $\beta=0$ is summarized in Table 2.
While the trapping horizon is absent in the region of $v<0$, there is a single future outer trapping horizon in the region of $v>0$.
This trapping horizon is spacelike and its area is increasing because $aa'>0$ holds.
According to the result in Sec.~\ref{Sec:THsub}, all the standard energy conditions are respected (violated) in the region of $v\le (>)0$ independent of the sign of $B_0$.
Figure~\ref{Fig-vr-plane-all} shows the $(r,v)$-plane for $\beta=0$.
It is seen that the region $v\ge 0$ represents an evolving black hole.
%------------<fig>---------------------------
\begin{figure}[htbp]
\begin{center}
%\rotatebox{-90}{
\includegraphics[width=0.4\linewidth]{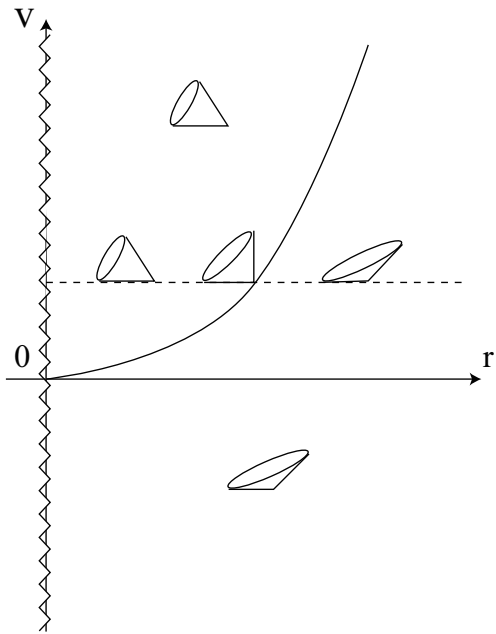}
%\subfigure[]{\includegraphics[width=0.7\linewidth]{Roberts-lambda1.eps}}
%\subfigure[]{\includegraphics[width=0.7\linewidth]{Roberts-lambda2.eps}}
%}
\caption{\label{Fig-vr-plane-beta0} $(r,v)$-plane for $\beta=0$ with $B_0\ne 0$.
A zigzag line and a solid curve represent a curvature singularity and a future outer trapping horizon, respectively.
Several future light cones are put to clarify the trapped region and the signature of the trapping horizon.}
\end{center}
\end{figure}
%--------------<fig>-----------------------

On the other hand, the behavior of the function $a$ governed by the master equation (\ref{master-a}) is quite complicated for $\beta\ne 0$ and depends on the sign of $B_0$ and $a_0$.
(See Table~\ref{table:signature} for the sign of $\beta$.)
%-------------- TABLE ---------------%
\begin{table}[htb]
\begin{center}
\caption{\label{table:signature} The sign of $a_0$ depending on $B_0$ and $\beta$.}
\begin{tabular}{|c||c|c|}
\hline \hline
 & $\beta>0$ & $\beta<0$    \\\hline
 $B_0>0$ & $a_0<0$ & $a_0>0$ \\\hline
$B_0<0$ & $a_0>0$ & $a_0<0$  \\ 
\hline \hline
\end{tabular}
\end{center}
\end{table} 
%------------------------------------%

The static solution of the master equation (\ref{master-a}) is given by Eq.~(\ref{a0-def}):
\begin{equation}
a_0=-\epsilon\biggl|\frac{8B_0 }{3 \kappa \beta}\biggl|^{1/3}.\label{a-static} 
\end{equation}
On the other hand, the dynamical solution is given by Eq.~(\ref{param}):
\begin{equation} 
\frac12\ln \left(\frac{(a-a_0)^2}{a^2+a_0 a+a_0^2}\right)-\sqrt{3} \arctan\left(\frac{2 a+a_0}{\sqrt{3} a_0}\right)=\frac{3 a_0^2 \beta }{4} (v-v_0),\label{solution-a1}
\end{equation}
where $v_0$ is an integration constant and $a_0$ is given by Eq.~(\ref{a-static}), namely the value of $a(v)$ for the static solution.
Hereafter, without loss of generality, we set 
\begin{equation} 
v_0=\frac{2\sqrt{3}\pi   }{ 9 \beta a_0^2} \label{v0-normal}
\end{equation}
by using the shift transformation of $v$ such that $a(0)=0$.
Then, the behavior of $a$ near $v=0$ is given by  
\begin{equation} 
a(v)\simeq \frac{2B_0}{3\kappa}v, \label{asymp-a0}
\end{equation}
which is the same as Eq.~(\ref{solution-a0}) for $\beta=0$.
It is noted that Eq.~(\ref{v0-normal}) gives $v_{\rm min}=0$ in Eq.~(\ref{v-minimum}).
Therefore, Xu's coordinate system (\ref{Xu-metric}) covers the region of $v>0$ in our coordinate system.
Also, the sign of $a$ in the metric function cannot change in his coordinate system as explained.
However, our coordinate system overcomes these two problems simultaneously.

\begin{figure}[htbp]
\begin{center}
%\rotatebox{-90}{
\includegraphics[width=1.0\linewidth]{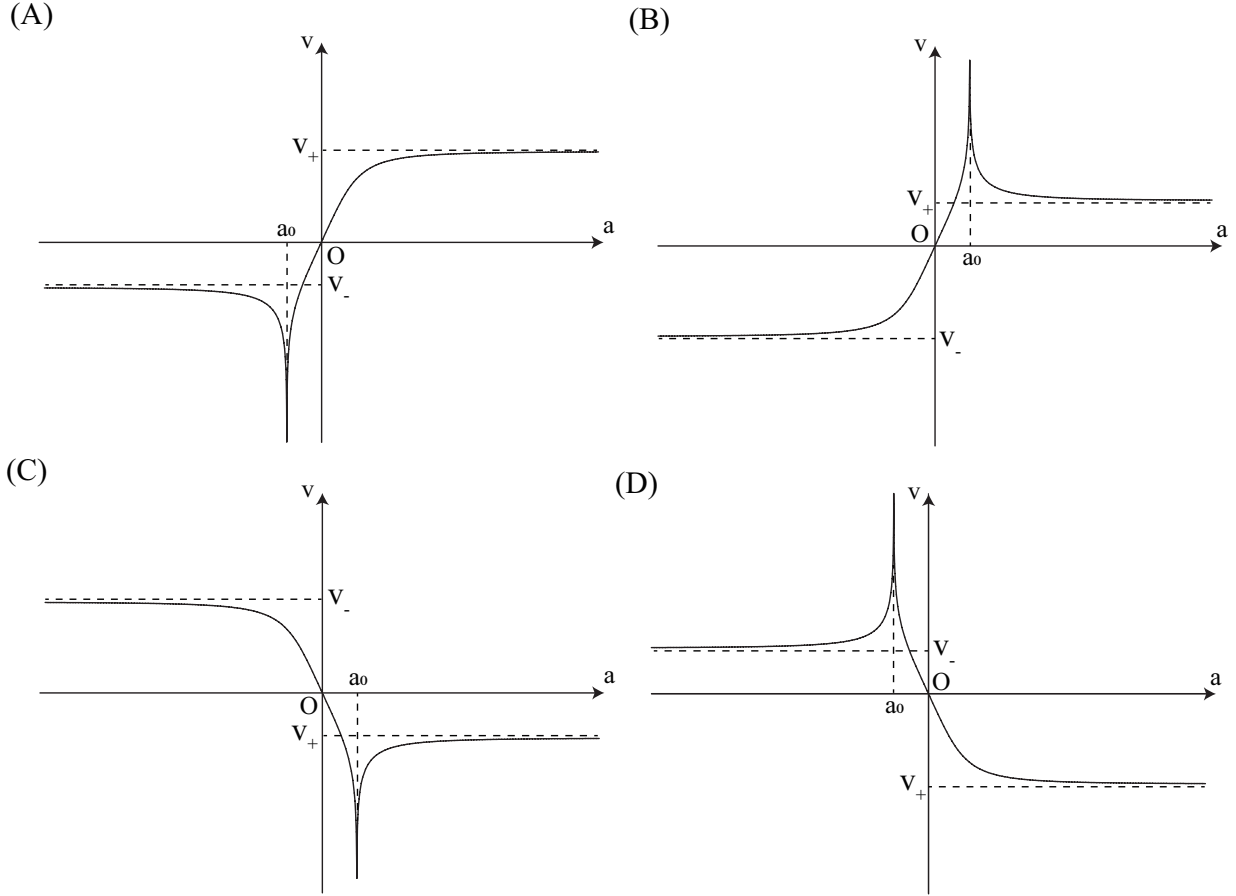}
%\subfigure[]{\includegraphics[width=0.7\linewidth]{Roberts-lambda1.eps}}
%\subfigure[]{\includegraphics[width=0.7\linewidth]{Roberts-lambda2.eps}}
%}
\caption{\label{Fig-v-B0a0ALL} The functions $v(a)$ given by Eq.~(\ref{solution-a1}) with Eq.~(\ref{v0-normal}) for (A) $B_0>0$ and $\beta>0$ (hence $a_0<0$), (B) $B_0>0$ and $\beta<0$ (hence $a_0>0$), (C) $B_0<0$ and $\beta>0$ (hence $a_0>0$), and (D) $B_0<0$ and $\beta<0$ (hence $a_0<0$).
$v_+$ and $v_-$ are defined by Eq.~(\ref{v+v-}).
The domains $a<a_0$ and $a>a_0$ represent distinct spacetimes because $a=a_0$ corresponds to $|v|\to \infty$.
 While there is no trapping horizon in the regions where $B_0a<0$ holds, there is a future outer trapping horizon in the regions where $B_0a>0$ holds.
The spacetimes (B) and (D) converge to a static black hole in far future, while the spacetimes (A) and (C) converge to a static naked singularity in far past.}
\end{center}
\end{figure}

The behavior of the functions $v(a)$, given by Eq.~(\ref{solution-a1}) with Eq.~(\ref{v0-normal}) are shown in Fig.~\ref{Fig-v-B0a0ALL}.
Figure~\ref{Fig-v-B0a0ALL} shows that the domains $a<a_0$ and $a>a_0$ represent distinct spacetimes because $a=a_0$ corresponds to $|v|\to \infty$.
Namely, there are two different branches of solutions in the dynamical solution (\ref{solution-a1}).

%------------<fig>---------------------------
%--------------<fig>-----------------------

The solution (\ref{solution-a1}) shows that $a(v)\to a_0$, namely the spacetime is asymptotically static, for $\beta v\to -\infty$.
Since the static solution represents a black hole (naked singularity) for $\beta<(>)0$ by Eq.~(\ref{BH-condition-static}), the dynamical solution is asymptotically static black hole in far future for $\beta< 0$.
In contrast, the solution is asymptotically static but naked singular solution in the far past for $\beta>0$.
On the other hand, for $\beta v\to +\infty$, $a(v)$ is not real and hence the solution is unphysical.

With Eq.~(\ref{v0-normal}), it is also shown that $a(v)$ diverges as $a\to \pm\infty$ for $v\to v_\pm$, where
\begin{equation} 
v_{\pm}:=v_0\mp\mbox{sign}{(a_0)}\frac{2\sqrt{3}\pi}{3\beta a_0^2}=v_0(1\mp 3\mbox{sign}{(a_0)}).
\end{equation}
More explicitly, $v_+$ and $v_-$ are given by 
\begin{eqnarray}
  v_+ = \begin{cases}
    -2v_0 & (\mbox{for}~~a_0>0) \\
   4v_0 & (\mbox{for}~~a_0<0)
  \end{cases}
,\qquad
  v_- = \begin{cases}
    4v_0 & (\mbox{for}~~a_0>0) \\
    -2v_0 & (\mbox{for}~~a_0<0)
  \end{cases}
. \label{v+v-}
\end{eqnarray}
By Eq.~(\ref{v0-normal}), $\beta$ determines the sign of $v_+$ and $v_-$.
$v=v_+$ and $v=v_-$ are curvature singularities with the following blow-up rate:
\begin{equation} 
a(v)^2\simeq-\frac{2}{\beta(v-v_\pm)} \to +\infty.  \label{a-v_s} 
\end{equation}
The above equation shows that $v<v_\pm$ and $v>v_\pm$ are physical regions with a real metric for $\beta>0$ and $\beta<0$, respectively.
It also shows that $\D v/\D a \to 0$ is realized for $v\to v_\pm$. 
Equation~(\ref{rh-infity}) shows that $r_{\rm h}\to \infty$ holds near these singularities if $B_0a>0$ is satisfied around there.

Lastly, $a(v)$ is shown to be monotonic because Eq.~(\ref{solution-a1}) shows
\begin{equation} 
\frac{\D a}{\D v}=\frac{\beta}{4}(a-a_0)(a^2+a_0a+a_0^2)\label{dadv}
\end{equation}
and $a(v)=a_0$ is not realized for any finite value of $v$.
Now all the information obtained up to now are summarized in Tables~\ref{table:domain-v1} and \ref{table:domain-v2} for $B_0>0$ and $B_0<0$, respectively.
There is a single future outer trapping horizon in the regions with T, while there is no trapping horizon in the regions with U.

%-------------- TABLE ---------------%
\begin{table}[htb]
\begin{center}
\caption{\label{table:domain-v1} The behavior of $a(v)$ for $B_0>0$. See the caption of Table~\ref{table:domain-v3}.}
\begin{tabular}{|c|c||c|c|c|c|c|c|c|c|c||c|c|}
\hline \hline
 $\beta$ & Branch & $v=-\infty$ & $\cdots$ & $v_-$ & $\cdots$ & $0$ & $\cdots$ & $v_+$ & $\cdots$ & $+\infty$ & Fig.~\ref{Fig-vr-plane-all}  \\\hline\hline
$+$ & $a<a_0$& $a_0(<0)$ & $\searrow$ (U) & $-\infty$ & n.a. & n.a. & n.a. & n.a. & n.a. & n.a. & (a)\\  \cline{2-12}
 & $a>a_0$ & $a_0(<0)$ & $\nearrow$ (U) & $\nearrow$ (U) & $\nearrow$ (U) & $0$ & $\nearrow$ (T) & $+\infty$ & n.a. & n.a. & (b) \\ \hline
$-$ & $a<a_0$ & n.a. & n.a. & $-\infty$ & $\nearrow$ (U) & 0 & $\nearrow$ (T) & $\nearrow$ (T) & $\nearrow$ (T) & $a_0(>0)$ & (c) \\  \cline{2-12}
 & $a>a_0$ & n.a. & n.a. & n.a. & n.a. & n.a. & n.a. & $+\infty$ & $\searrow$ (T) & $a_0(>0)$ &(d) \\ 
\hline \hline
\end{tabular}
\caption{\label{table:domain-v2} The behavior of $a(v)$ for $B_0<0$. See the caption of Table~\ref{table:domain-v3}.}
\begin{tabular}{|c|c||c|c|c|c|c|c|c|c|c||c|c|}
\hline \hline
 $\beta$ & Branch & $v=-\infty$ & $\cdots$ & $v_+$ & $\cdots$ & $0$ & $\cdots$ & $v_-$ & $\cdots$ & $+\infty$ & Fig.~\ref{Fig-vr-plane-all}   \\\hline\hline
$+$ & $a<a_0$ & $a_0(>0)$ & $\searrow$ (U) & $\searrow$ (U) & $\searrow$ (U) & $0$ & $\searrow$ (T) & $-\infty$ & n.a. & n.a. & (b) \\  \cline{2-12}
 & $a>a_0$ & $a_0(>0)$ & $\nearrow$ (U) & $+\infty$ & n.a. & n.a. & n.a. & n.a. & n.a. & n.a. & (a) \\ \hline
$-$ & $a<a_0$ & n.a. & n.a. & n.a. & n.a. & n.a. & n.a. & $-\infty$ & $\nearrow$ (T) & $a_0(<0)$ & (d) \\  \cline{2-12}
 & $a>a_0$ & n.a. & n.a. & $+\infty$ & $\searrow$ (U) & 0 & $\searrow$ (T) & $\searrow$ (T) & $\searrow$ (T) & $a_0(<0)$ & (c) \\ 
\hline \hline
\end{tabular}
\end{center}
\end{table} 
%------------------------------------%

%------------<fig>---------------------------
\begin{table*}[htb]
\begin{center}
\caption{ The corresponding $(r,v)$-plane in Fig.~\ref{Fig-vr-plane-all} depending on the parameters, where the values of $v_{\rm s}$ in the figure are also shown. }
\label{table:vr}
\begin{tabular}{|c|c|c|c|}\hline
& & Branch $a<a_0$  & Branch $a>a_0$  \\ \hline
$B_0>0$& $\beta > 0$ &  (a) ($v_{\rm s}=v_-$)  &  (b) ($v_{\rm s}=v_+$)   \\ \cline{2-4}
 & $\beta< 0$ &  (c) ($v_{\rm s}=v_-$)  &  (d) ($v_{\rm s}=v_+$)  \\ \hline
$B_0<0$ & $\beta> 0$ & (b) ($v_{\rm s}=v_-$)  &  (a) ($v_{\rm s}=v_+$) \\ \cline{2-4}
& $\beta< 0$ & (d) ($v_{\rm s}=v_-$)  &  (c) ($v_{\rm s}=v_+$)  \\ \hline
\end{tabular}
\end{center}
\end{table*}
%------------<fig>---------------------------

The $(r,v)$-planes for $\beta\ne 0$ with $B_0\ne 0$ are presented in Fig.~\ref{Fig-vr-plane-all}.
Each $(r,v)$-plane is divided into two portions by a curvature singularity $v=v_+$ or $v=v_-$ and the function $a(v)$ is real only in one portion.
The correspondence between the panels in Fig.~\ref{Fig-vr-plane-all} and the parameters in the solution is shown in Table~\ref{table:vr} as well as in Tables~\ref{table:domain-v1} and \ref{table:domain-v2}.
In Fig.~\ref{Fig-vr-plane-all}, according to the result in Sec.~\ref{sec:energy}, all the standard energy conditions are violated everywhere in (a) and (d) and in the region of $v>0$ in (b) and (c).
On the other hand, NEC, WEC, and SEC are respected in the region of $v\le 0$ in (b) and (c).
%------------<fig>---------------------------
%--------------<fig>-----------------------

\subsubsection{Structure of the singularity at $v=v_\pm$}

We have already shown that the singularity $r=0$ with $v\ne v_\pm$ in the trapped (untrapped) region is spacelike (timelike).
In order to identify the Penrose diagrams corresponding the panels (a)--(d) in Fig.~\ref{Fig-vr-plane-all}, one has to clarify the structure of the singularity at $v=v_\pm$, which appears only for $\beta\ne 0$. 
For this purpose, we use the following lemmas.
\begin{lm}
\label{lm:singularity1}
In the panels (a)--(d) in Fig.~\ref{Fig-vr-plane-all}, there is no future-directed outgoing radial null geodesic emanating from or terminating at $v=v_{\pm}$ with a finite positive value of $r$.
\end{lm}
{\it Proof:}
A future-directed outgoing radial null geodesic $\gamma$, which is given by $r=r_\gamma(v)$ satisfies Eq.~(\ref{ODE-radialnull}), namely
\begin{align}
\frac{\D r_\gamma}{\D v}=\frac12\biggl(\frac{r_\gamma^2}{l^2}-B_0 a(v)-\frac{B_0 \kappa  a(v)^3}{12r_\gamma}\biggl). \label{FDRDG}
\end{align}
Suppose that $\gamma$ emanates from or terminates at $v=v_{\pm}$ with a finite positive value of $r(=r_0)$.
Evaluating Eq.~(\ref{FDRDG}) at $v=v_{\pm}$ with $r=r_0$, where $a(v)$ behaves as Eq.~(\ref{a-v_s}), we obtain 
\begin{align}
\frac{\D r_\gamma}{\D v}\propto |v-v_{\pm}|^{-3/2}.
\end{align}
This gives a contradiction since $r_\gamma$ blows up for $v\to v_{\pm}$ as $|r_\gamma|\propto |v-v_{\pm}|^{-1/2}\to \infty$ and hence no future-directed outgoing radial null geodesic emanates from or terminates at $v=v_{\pm}$ with a finite positive value of $r$.
\qed

\begin{lm}
\label{lm:singularity2}
In the panels (a) and (d) in Fig.~\ref{Fig-vr-plane-all}, there is an infinite number of future-directed outgoing radial null geodesics satisfying $r\to \infty$ for $v\to v_\pm$ and there is no such geodesic satisfying $r\to 0$ for $v\to v_\pm$.
In Figs.~\ref{Fig-vr-plane-all}(b) and \ref{Fig-vr-plane-all}(c), there is an infinite number of future-directed outgoing radial null geodesics satisfying $r\to 0$ for $v\to v_\pm$ and there is no such geodesic satisfying $r\to \infty$ for $v\to v_\pm$.
\end{lm}
{\it Proof:}
A small neighbourhood of $v=v_{\pm}$ with a finite positive $r$ is denoted by ${\cal D}$.
Since the Lipschitz condition for the ordinary differential equation (\ref{FDRDG}) is satisfied, there exists a future-directed outgoing radial null geodesic $\gamma$ with an initial condition in ${\cal D}$.

As shown by the future light-cones in Fig.~\ref{Fig-vr-plane-all}, $\D r_\gamma/\D v>0$ holds near $v=v_{\pm}$ in (a) and (c).
Since the metric function $f$ blows up for $v\to v_{\pm}$, Eq.~(\ref{ODE-radialnull}) shows that light cones in (a) and (c) open completely for $v\to v_{\pm}$.
Then, by Lemma~\ref{lm:singularity1}, we have $r\to \infty$ in (a) and $r\to 0$ in (c) in the limit of $v\to v_{\pm}$ along $\gamma$.

On the other hand, the future light-cones in Fig.~\ref{Fig-vr-plane-all} show that $\D r_\gamma/\D v<0$ holds near $v=v_{\pm}$ in (b) and (d).
For $v\to v_{\pm}$, light cones in (b) and (d) close completely.
Then, by Lemma~\ref{lm:singularity1}, we have $r\to 0$ in (b) and $r\to \infty$ in (d) in the limit of $v\to v_{\pm}$ along $\gamma$.

For a given $\gamma$, there is always a finite interval between $v=v_\pm$ and $\gamma$ in the $(r,v)$-plane.
Any future-directed outgoing radial null geodesic starting from a point in this interval cannot intersect with $\gamma$ because the Lipschitz condition is satisfied.
Hence it necessarily arrives $r=0$ or $r \to \infty$ for $v\to v_\pm$ by Lemma~\ref{lm:singularity1}.
Since this interval is a finite domain, there exists an infinite number of such geodesics and therefore this portion of the singularity at $v=v_{\pm}$ is ingoing null.
\qed

%------------<fig>---------------------------
\begin{table*}[htb]
\begin{center}
\caption{Structures of $v=v_\pm$ in the Penrose diagrams for the panels (a)--(d) in Fig.~\ref{Fig-vr-plane-all}.}
\label{table:structure}
\begin{tabular}{|c|c|c|c|}\hline
  Panel    & $r=0$ & $r\in(0,\infty)$ & $r\to \infty$ \\ \hline\hline
(a)  & point & point  & ingoing null   \\ \hline
(b)  & ingoing null & point & point  \\ \hline
(c) & ingoing null & point  & point \\  \hline
(d) & point & point  & ingoing null  \\ \hline
\end{tabular}
\end{center}
\end{table*}
%------------<fig>---------------------------

By Lemmas~\ref{lm:singularity1} and \ref{lm:singularity2}, the structures of $v=v_\pm$ in the Penrose diagrams are summarized in Table~\ref{table:structure}.
Actually, the point $v=v_\pm$ with $r\in[0,\infty)$ is a curvature singularity because the Kretschmann invariant (\ref{K-sol}) blows up for $a\to \pm\infty$ with a finite $r$.
On the other hand, the point $v=v_\pm$ with $r\to \infty$ is at least a p.p. curvature singularity by the following lemma.
\begin{lm}
\label{lm:singularity3}
For $\beta\ne 0$, $a^2/r_{\gamma}\to \infty$ holds for $a\to \pm\infty$.
\end{lm}
{\it Proof:}
Putting $r_{\gamma}(v)=a(v)^2\eta(v)$ in Eq.~(\ref{FDRDG}), we obtain
\begin{align}
\frac{1}{12\kappa}(3\beta\kappa a^3+8B_0)\biggl(2\eta+a\frac{\D\eta}{\D a}\biggl)=\frac12\biggl(\frac{a^3\eta^2}{l^2}-B_0-\frac{B_0 \kappa}{12\eta}\biggl), \label{eta-eval}
\end{align}
where we used $\eta'=a'\D \eta/\D a$ and Eq.~(\ref{master-a}).
Suppose $\eta$ blows up or converges to a nonzero constant for $a\to \pm\infty$. Then Eq.~(\ref{eta-eval}) is approximated to be 
\begin{align}
\beta\biggl(2\eta+a\frac{\D\eta}{\D a}\biggl)\simeq\frac{2\eta^2}{l^2}
\end{align}
near $a\to \pm\infty$.
This is an ordinary differential equation for $\eta(a)$ and its solution is 
\begin{align}
\eta(a)=\frac{\beta l^2}{1+\eta_0 \beta l^2 a^2},
\end{align}
where $\eta_0$ is an integration constant.
This solution gives a contradiction $\lim_{a\to \pm\infty}\eta\to 0$.
Thus, $\eta\to 0$ for $a\to \pm\infty$ is concluded.
\qed

By Lemma~\ref{lm:singularity3}, the component of the Riemann tensor $R_{(0)(2)(1)(2)}$ in the orthonormal frame given by Eq.~(\ref{R-ortho}) blows up for $v\to v_\pm$ (and hence $a\to \pm\infty$):
\begin{align}
|R_{(0)(2)(1)(2)}|=&\biggl|\frac{3\kappa\beta^2a_0^3(a-a_0)(a^2+a_0a+a_0^2)(\kappa a^2+4r)}{512r^2}\biggl|\to \infty,
\end{align}
where we used Eqs.~(\ref{B0-beta}) and (\ref{master-a}).
Therefore, $v=v_\pm$ with $r\to \infty$ is a curvature singularity.

\begin{figure}[htbp]
\begin{center}
%\rotatebox{-90}{
\includegraphics[width=0.7\linewidth]{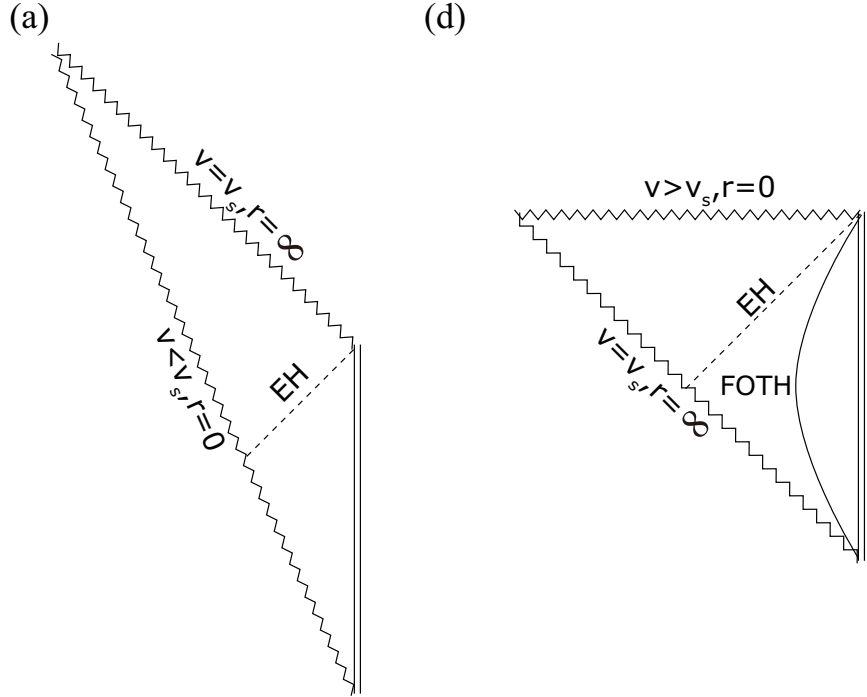}
%\subfigure[]{\includegraphics[width=0.7\linewidth]{Roberts-lambda1.eps}}
%\subfigure[]{\includegraphics[width=0.7\linewidth]{Roberts-lambda2.eps}}
%}
\caption{\label{Fig-Penrose-all3} The Penrose diagrams for the $(r,v)$-planes presented by the panels (a) and (d) in Fig.~\ref{Fig-vr-plane-all}.
EH and FOTH stand for the event horizon and a future outer trapping horizon, respectively.
A double line is the AdS infinity.
In (a), the spacetime is asymptotically static naked-singular solution in far past.
In spite that the spacetime is everywhere untrapped, there is an event horizon in (a). 
In (d), the spacetime is asymptotically static black hole in far future.
The trapping horizon is timelike and the spacetime represents a shrinking black hole.}
\end{center}
\end{figure}

The Penrose diagrams for the panels (a) and (d) in Fig.~\ref{Fig-vr-plane-all} are drawn in Fig.~\ref{Fig-Penrose-all3}.
In both cases, all the standard energy conditions are violated everywhere.
While the diagram (d) represents a shrinking black hole, the diagram (a) admits an event horizon without any trapped surface in the spacetime.
Hence, the diagram (a) represents a black hole in the sense of the event horizon but it is not a black hole in the sense of a future outer trapping horizon.
This curious configuration should be a consequence of the violation of the null energy condition.

The panels (c) and (d) in Fig.~\ref{Fig-vr-plane-all} show that the spacetime is future asymptotically HMTZ black hole \cite{Henneaux:2002wm}.
However, it was shown that, for $\alpha=0$, HMTZ black hole is dynamically unstable against radial linear perturbations~\cite{martinez1998}.
This implies that (i) the present solution with $\alpha=0$ does not contain the unstable modes found in~\cite{martinez1998}, (ii) the boundary conditions at $r\to \infty$ for the present solution and the perturbations in~\cite{martinez1998}  are different, or (iii) the HMTZ black hole with $\alpha=0$ is nonlinearly stable. 
In the next subsection, we will focus on the $(r,v)$-planes (b) and (c) in Fig.~\ref{Fig-vr-plane-all}.
%------------<fig>---------------------------
%--------------<fig>-----------------------

\subsection{Gravitational collapse from regular initial data}

\subsubsection{Attachment at $v=0$ to the massless BTZ spacetime}
We have seen that there is a naked singularity in the region of $v<0$ in Fig.~\ref{Fig-vr-plane-beta0} for $\beta=0$ and in the panels (b) and (c) in Fig.~\ref{Fig-vr-plane-all} for $\beta\ne 0$.
In order to construct a spacetime representing the formation of a black hole, we take our solution (\ref{sol-attach})--(\ref{def-phi0}) in the region of $v\ge 0$, where all the standard energy conditions are violated, and attach it to the past massless BTZ spacetime with a vanishing scalar field $\phi\equiv 0$ for $v\le 0$ at a null hypersurface $v=0$, which we denote $\Sigma$.
Such a massless BTZ solution for $v\le 0$ is realized by Eqs.~(\ref{sol-attach})--(\ref{def-phi0}) with $a(v)\equiv 0$.

Actually, we can show that the matching null hypersurface $\Sigma$ is regular, namely there is no massive thin-shell at $v=0$.
In general relativity, as seen in Sec.~\ref{sec:practice} for a null dust fluid, continuity of the induced metric $\sigma_{AB}$ and the transverse curvature $C_{ab}$ at the matching null hypersurface $\Sigma$ are sufficient for the the absence of a massive thin-shell.
In the present system, in contrast, since the field equations (\ref{Einstein}) contain second derivative of the scalar field $\nabla_\mu\nabla_\nu\phi$, the jump of the first derivative of the scalar field also contributes to the energy-momentum tensor on $\Sigma$.
As a result, continuity of $\sigma_{AB}$, $C_{ab}$, $\phi$, and $\phi (N^\mu\nabla_\mu\phi)$ at $\Sigma$ are sufficient for the absence of a massive thin-shell in the present system~\cite{forthcoming}.

$\sigma_{AB}$ and $C_{ab}$ for a null hypersurface $v=0$ in the spacetime (\ref{metric-assumption}) are given by Eqs.~(\ref{hab}) and (\ref{trans-C0}).
Because our solution (\ref{sol-attach})--(\ref{def-phi0}) shows $f(0,r)=r^2/l^2$ and $\phi(0,r)=0$, $\sigma_{AB}$, $C_{ab}$, and $\phi$ are continuous at $\Sigma$.

Now let us see  $\phi(N^\mu\nabla_\mu\phi)$ at $\Sigma$.
Using Eqs.~(\ref{N-attachment}) and (\ref{def-phi0}), we compute
\begin{align}
(N^\mu\nabla_\mu\phi)|_{\Sigma}=&\frac{1}{4(r+\kappa a(0)^2/8)^{3/2}}\biggl\{r\biggl(\beta a(0)^3+\frac{8 B_0}{3 \kappa }\biggl)-a(0)f(0,r)\biggl\} \nonumber \\
=&\frac{2 B_0}{3 \kappa r^{1/2}}, \label{phi-jump}
\end{align}
where we used Eq.~(\ref{master-a}), $a(0)=0$, and $f(0,r)=r^2/l^2$.
Since $\phi(0,r)=0$ holds in our non-stealth solution, we obtain
\begin{align} \label{normalderphi}
(\phi N^\mu\nabla_\mu\phi)|_{\Sigma}=0.
\end{align}
Because  Eq. \eqref{normalderphi}  also holds in the past massless BTZ spacetime with a vanishing scalar field, $\phi(N^\mu\nabla_\mu\phi)$ is continuous at $\Sigma$.
Therefore, the energy-momentum tensor on $\Sigma$ is vanishing, namely there is no massive thin-shell there. Moreover, since $\phi(0,r)=0$ the field equation \eqref{FE} is continuous on $\Sigma$.

One might consider the past massless BTZ spacetime with a stealth scalar field instead of a vanishing scalar field for $v\le 0$.
Such stealth scalar fields are given by Eqs.~(\ref{stealth-scalar1-3}), (\ref{stealth-scalar1-4}), (\ref{53}), or (\ref{pb}).
However, the first three do not satisfy $\phi=0$ with constant $v$ and therefore $\phi$ cannot be continuous at $\Sigma$.
On the other hand, $\phi=0$ is possible at $\Sigma$ in the last case (\ref{pb}) with $a_0=0$ but this reduces to the case of a vanishing scalar field.
Therefore, when we consider the massless BTZ spacetime  with a stealth scalar field for attachment, $\Sigma$ cannot be regular.

\subsubsection{Nature of the singularity at $r=v=0$}
\label{Sec:r=v=0}
We have shown that our solution (\ref{sol-attach})--(\ref{def-phi0}) for $v\ge 0$ can be attached to the past massless BTZ spacetime for $v\le 0$ without a massive thin-shell at $v=0$.
In the region of $v\ge 0$, there is a future outer trapping horizon $r=r_{\rm h}(v)$ which is an increasing function with $r_{\rm h}(0)=0$.
(See Fig.~\ref{Fig-vr-plane-beta0} for $\beta=0$ and the panels (b) and (c) in Fig.~\ref{Fig-vr-plane-all} for $\beta\ne 0$.)
We express its inverse as $v=v_{\rm TH}(r)$ which is also monotonic and satisfies $v_{\rm TH}(0)=0$. 
Now we study the nature of the singularity at $r=v=0$, where $a(v)$ behaves as Eq.~(\ref{asymp-a0}).

By Eqs.~(\ref{a-r-TH}) and (\ref{asymp-a0}), the behavior of the trapping horizon $v=v_{\rm TH}(r)$ near $r=0$ is given by 
\begin{eqnarray}
v_{\rm TH}\simeq \frac{3\kappa}{2B_0^2l^2}r^2 \label{v-r-TH}
\end{eqnarray}
for any $\beta$.
Because this is an increasing function of $r$, there may exist future-directed outgoing causal geodesics emanating from the singularity at $v=r=0$.
Such causal geodesics $v=v_{\rm CG}(r)$ satisfies $v_{\rm CG}(r)<v_{\rm TH}(r)$ near $v=r=0$ because they  cannot enter the trapped region. 
Let us check whether there are such geodesics or not.

By the contraposition of Lemma~\ref{lm:geodesics}, it is sufficient to prove the absence of future-directed outgoing radial null geodesics to conclude that the singularity is censored.
Actually, the singularity at $v=r=0$ is censored by the following lemma.

\begin{lm}
\label{lm:singularity-main}
The singularities at $v=r=0$ in Fig.~\ref{Fig-vr-plane-beta0} (for $\beta=0$ with $B_0\ne 0$) and in the panels (b) and (c) in Fig.~\ref{Fig-vr-plane-all} (for $\beta \ne 0$ with $B_0\ne 0$) are censored.
\end{lm}
{\it Proof:}
Using Eq.~(\ref{master-a}), we write the future-directed radial null geodesic equation (\ref{FDRDG}) for $r_\gamma=r_\gamma(a)$ as
\begin{align}
\frac{1}{12\kappa}(3 \beta  \kappa  a^3+8 B_0)\frac{\D r_\gamma}{\D a}=\frac12\biggl(\frac{r_\gamma^2}{l^2}-B_0 a-\frac{B_0 \kappa  a^3}{12r_\gamma}\biggl).\label{FDRDG-modified1}
\end{align}
We have adopted Eq.~(\ref{v0-normal}) so that $a(0)=0$ holds.
Because the three terms in the bracket in the right-hand side cannot be of the same order simultaneously, at least one of those three is negligible in the vicinity of $a=r=0$. 
Among them, the second term cannot be negligible because this condition ($a\ll r_\gamma^2$ and $a\ll a^3/r_\gamma$) gives $|a|^{1/2}\ll r_\gamma\ll a^2$, which is not satisfied for $a\to 0$.
Thus, there are three possibilities around $r=a=0$: (i) only the second term dominates, (ii) the first and the second terms dominate with the same order, and (iii) the second and the third terms dominate with the same order.

In the case (i), Eq.~(\ref{FDRDG-modified1}) is approximated around $a=r=0$ by $\D r_\gamma/\D a=-3\kappa a/4$, which is integrated to give $r_\gamma=-3\kappa a^2/2+r_0$, where $r_0$ is an integration constant.
This case (i) is discarded because $r_0=0$ is required for $r_\gamma(0)=0$ but then $r_\gamma(a)$ is negative near $a=0$.

The case (ii) implies that $r_\gamma\propto |a|^{1/2}$ holds and Eq.~(\ref{FDRDG-modified1}) is approximated by 
\begin{align}
\frac{\D r_\gamma}{\D a}\simeq \frac{3\kappa}{4B_0}\biggl(\frac{r_\gamma^2}{l^2}-B_0 a\biggl)\label{FDRDG-modified2}
\end{align}
around $a=r=0$.
However, the above equation is not satisfied because $r_\gamma\propto |a|^{1/2}$ gives $\D r_\gamma/\D a\propto |a|^{-1/2}$ and hence the case (ii) is also discarded. 

The case (iii) implies that $r_\gamma\simeq r_1 a^2$ holds, where $r_1$ is a positive constant, and Eq.~(\ref{FDRDG-modified1}) is approximated by 
\begin{align}
\frac{\D r_\gamma}{\D a}\simeq -\frac{3\kappa}{4}\biggl(a+\frac{\kappa  a^3}{12r_\gamma}\biggl).\label{FDRDG-modified3}
\end{align}
Putting $r_\gamma\simeq r_1 a^2$ into Eq.~(\ref{FDRDG-modified3}), we obtain $(4r_1+\kappa)(8r_1+\kappa)=0$, which show that $r_1$ is negative. 
Hence, the case (iii) is also discarded. 

Because there is no future-directed radial null geodesic emanating from $r=a=0$ in all the cases, the singularity at $r=v=0$ is censored.
\qed

\begin{figure}[htbp]
\begin{center}
%\rotatebox{-90}{
\includegraphics[width=0.8\linewidth]{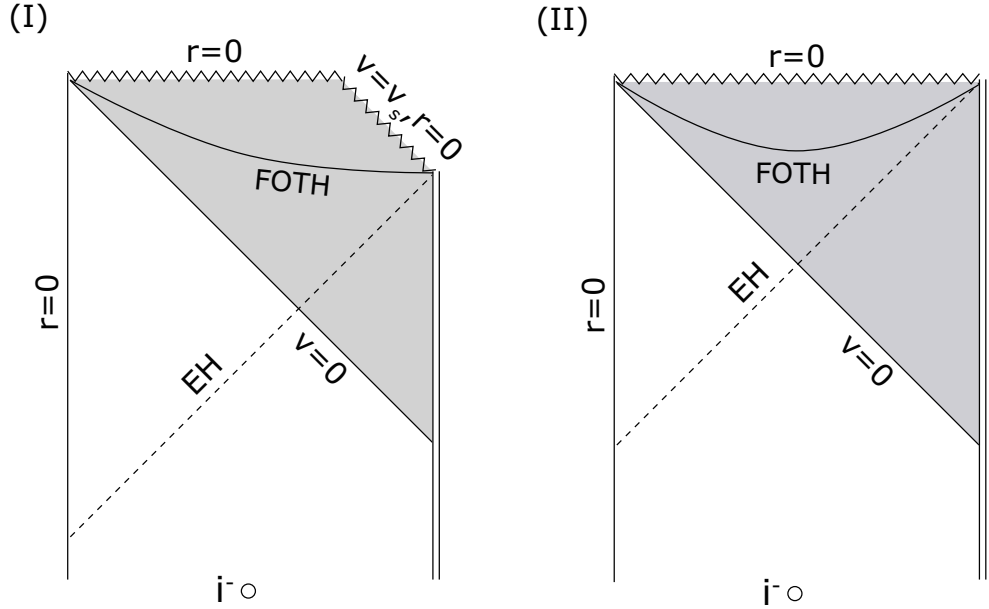}
%\subfigure[]{\includegraphics[width=0.7\linewidth]{Roberts-lambda1.eps}}
%\subfigure[]{\includegraphics[width=0.7\linewidth]{Roberts-lambda2.eps}}
%}
\caption{\label{Fig-Penrose-all-BH} (I) The Penrose diagram for the solution with $\beta > 0$ given by the panel (b) in Fig.~\ref{Fig-vr-plane-all}, attached to the past massless BTZ spacetime at $v=0$. (The scalar field is nontrivial in the shadowed region.)
(II) The Penrose diagram for the solution with $\beta=0$ given by Fig.~\ref{Fig-vr-plane-beta0} and the solution for $\beta < 0$ given by the panel (c) in Fig.~\ref{Fig-vr-plane-all}, attached to the past massless BTZ spacetime at $v=0$.
EH and FOTH stand for the event horizon and a future outer trapping horizon, respectively.
A double line is the AdS infinity, while $i^-$ denotes the past timelike infinity.}
\end{center}
\end{figure}

Collecting all the information obtained up to now, we can draw the Penrose diagrams for the resulting spacetimes as shown in Fig.~\ref{Fig-Penrose-all-BH}.
Both of them represent the black-hole formation in an asymptotically AdS spacetime from regular initial data.
%------------<fig>---------------------------
%--------------<fig>-----------------------

%

\section{Conclusions}
In the present paper, we have obtained two classes of exact dynamical and inhomogeneous solutions in three-dimensional AdS gravity with a conformally coupled scalar field.
The first class represents a stealth scalar field overflying the BTZ spacetime (\ref{sol-attach}) with $f(v,r)=-M_0+r^2/l^2$.
In this first class, the scalar field is given by Eqs.~(\ref{stealth-scalar1-3}), (\ref{stealth-scalar1-4}), (\ref{53}), or (\ref{pb}).
The other class is an asymptotically AdS solution with a non-stealth scalar field and explicitly given by Eqs.~(\ref{sol3f})--(\ref{sol3p}), where the function $a(v)$ is given by Eqs.~(\ref{param-beta=0}) and (\ref{param}) for $\beta=0$ and $\beta\ne 0$, respectively.
This non-stealth solution is an analytic extension of the solution obtained by Xu~\cite{Xu:2014xqa}.
Introducing a simpler coordinate system, we have found a new branch of solutions hidden in the Xu's original coordinate system.
We have investigated geometrical and physical properties of all the branches in detail and finally found that this solution represents a variety of physically interesting spacetimes depending on the parameters.

The solution for $\beta > 0$ given by the panel (a) in Fig.~\ref{Fig-vr-plane-all} represents a curious spacetime admitting an event horizon without any trapped surface.
In spite that this spacetime represents a dynamical black hole defined by the event horizon, a future outer trapping horizon, an alternative quasi-local definition of a black hole, is absent.
On the other hand, the solution for $\beta < 0$ given by the panel (d) in Fig.~\ref{Fig-vr-plane-all} represents an eternally shrinking dynamical black hole.
In both cases, all the standard energy conditions are violated in the whole spacetime.

Lastly, the solution for $\beta=0$ given by Fig.~\ref{Fig-vr-plane-beta0} and the solution for $\beta\ne 0$ given by the panels (b) and (c) in Fig.~\ref{Fig-vr-plane-all} can describe the black-hole formation in an asymptotically AdS spacetime from regular initial data, by attaching the solution at $v=0$ to the past massless BTZ spacetime with a vanishing scalar field in a regular manner.
Among them, the spacetime converges to a static HMTZ black hole given by Eq.~(\ref{HMTZ}) with Eq.~(\ref{statold}) in far future in the case of panel (c) in Fig.~\ref{Fig-vr-plane-all}.
Although the scalar field in the resulting spacetime violates all the standard energy conditions, this could be an interesting model to investigate the AdS${}_3$/CFT${}_2$ duality in a dynamical setting.

Undoubtedly, generalization of the present solutions into higher dimensions is quite interesting but it is a highly nontrivial task.
Such solutions could shed light on rich properties of asymptotically AdS spacetimes and expose their difference from the spacetimes with a vanishing or positive cosmological constant.
We leave this problem for future investigations.

\subsection*{Acknowledgements}
HM thanks Steven Willison and Masato Nozawa for valuable comments. 
HM also thanks the Theoretical Physics group in CECs and Universidad Adolfo Ib\'a\~{n}ez for hospitality and support, where a large part of this work was carried out. LA thanks the Conicyt grant 21160827. 
This work has been partially funded by the Fondecyt
grants  1161311 and 1180368. The Centro de Estudios Cient\'{\i}ficos (CECs) is funded by the Chilean Government through the Centers of Excellence Base Financing Program of Conicyt.

%\appendix

\end{document}